%% file: communication.tex
\documentclass[a4paper]{article}

\usepackage{graphicx}
\usepackage{subfigure}
\usepackage[round, sort&compress]{natbib}
\usepackage{authblk}

\include{communicationmacros}

\bibpunct{(}{)}{;}{a}{}{;}

\title{Expectation-driven interaction: 
	a model based on Luhmann's contingency approach.}

\author[1]{M.~J.~Barber}
\author[2]{Ph.~Blanchard}
\author[3]{E.~Buchinger}
\author[4]{B.~Cessac}
\author[1,2]{L.~Streit}
\affil[1]{Universidade da Madeira,
		Centro de Ci\^encias Matem\'aticas,
		Campus Universit\'ario da Penteada,
		9000-390 Funchal,
		Portugal}
\affil[2]{Bielefeld University,
		Faculty of Physics and BiBoS,
		Universit\"atsstr.~25,
		33615 Bielefeld,
		Germany}
\affil[3]{ARC Systems Research GmbH,
		A-2444  Seibersdorf, 
		Austria}
\affil[4]{Institut Non Lin\'eaire de Nice et Universit\'e de Nice,
		1361 Route des Lucioles,
		Valbonne 06560,
		France}

\begin{document}
\maketitle

\begin{abstract}
    We introduce an agent-based model of interaction, drawing on the contingency approach from Luhmann's theory of social systems. 
The agent interactions are 
defined by the exchange of distinct messages. Message selection is based on the history of 
the interaction and developed within the confines of the problem of double 
contingency. We examine interaction strategies in the light
of the message-exchange description using analytical and computational methods. 
\end{abstract}

\section{Introduction.} \label{sec:intro}
Contingency is a key component of the theory of social systems elaborated by the German sociologist Niklas Luhmann. The theory describes a vision of society as a self-reproducing (autopoietic) system of communicative interaction. It is based on the work of Talcott Parsons (sociology), Edmund Husserl (philosophy), Heinz von Foerster (cybernetics), Humberto R. Maturana/Francisco Varela (evolutionary biology), and George Spencer Brown (mathematics). The first 
comprehensive elaboration of the theory can be dated to the appearance of \booktitle{Soziale 
Systeme} (``Social Systems'') in \citeyear{Luh:1984}, with an English translation available 
in \citeyear{Luh:1995}, and found a kind of finalization with the appearance of \booktitle{Die 
Gesellschaft der Gesellschaft} (``The Society's Society'') in 1997. The most relevant texts for the 
modeling of learning as a social process can be found in \booktitle{Social Systems} chapter 4, 
``Communication and Action,'' (p.137--175) and in chapter 8, 
``Structure and Time'' (p.278--356). 
For readers who are interested in an exposition of Luhmann's ideas we recommend for a start 
\booktitle{Einf\"uhrung in die Systemtheorie} \citep{Luh:2004} and the 
February, \citeyear{Arn:2001} issue
of  \booktitle{Theory, Culture and Society},  particularly the 
contributions ``Niklas Luhmann: An 
introduction'' by Jakob \citeauthor{Arn:2001} and 
``Why systems?'' by Dirk \citeauthor{Bae:2001}.

The present paper is the first step of a common interdisciplinary work involving a sociologist
and theoretical physicists. Our ultimate goal is to design a multi-agent model
using the theoretical background of Luhmann's theory and based on a mathematical formalization of a social process in which expectation-driven communication results in learning.
We especially want to explore  knowledge diffusion and interactive
knowledge generation (innovation)
in various social structures, represented essentially as dynamically evolving graphs.
Primarily, we adopt a general point of view of modeling from theoretical physics and only secondarily  intend to have an ``as realistic as possible'' model. Rather than putting ``everything'' in, the focus is to distinguish a few features in Luhmann's theory that are 
relevant for our purposes, and propose a model of expectation-driven interaction that is a possibly rough first step, but is tractable either on an analytical or computational level. In particular, it is promising to use an approach based on dynamical systems theory and statistical mechanics to analyze models of social interactions, 
and there is some research activity in this field
\citep{StaHohPit:2004,ForSta:2005,Sta:2003,WeiDefAmb:2004,Wei:2004}.
On the one hand, dynamical systems theory provides mathematical tools to study the 
dynamical evolution of interacting agents at the ``microscopic'' level (specifically, the detailed evolution of each agent is considered). On the other, statistical physics allows in principle a description at the mesoscopic level (groups of agents) and macroscopic level (the population as a whole). 

However, though statistical physics gives accurate description of models in physics, one must attend to the  fact that interactions between human beings are  more complex than interactions usually considered in physics: they are non-symmetric, nonlinear, involve memory effects, are not explicitly determined by an energy function, \etc{} 
The present paper is an attempt to consider social mechanisms beyond state-of-the-art
modeling.

We introduce a mathematical model portraying interaction as an alternating exchange of messages by a pair of agents
where memories of the past exchanges are used to build 
evolving reciprocal
expectations. The resulting ``interaction sequence'' is quite rich and complex.
In fact, the
pairwise interactions between agents depend on various
independent parameters, and tuning them can lead to drastic changes in the model
properties. In  this work, we give some results in this direction, based on mathematical as well as computational
investigations.
These results are a necessary foundation for handling the intended multi-agent case.

The present model is a generalization of the work of Peter Dittrich, Thomas Krohn and Wolfgang Banzhaf \citeyearpar{DitKroBan:2003}. Based on Luhmann's theory of social systems, they designed a model  which describes the emergence of social order. We follow them in portraying communicative interaction as an alternating exchange of messages by a pair of agents. We use the memory structures they defined as a starting point for our agents. The memories are the key to the formalization of the structure of expectation in their model and we take the same approach. However, we introduce an additional form of memory to allow  individual variation in the expectations of the agents.

The main body of the paper is divided into four sections. First the conceptual background is elaborated, followed by the mathematical definition of the model. Next, the main mathematical and simulation results are discussed. The text ends with an outlook on planned extensions of the model.

\section{Conceptual background.}\label{sec:conceptualbackground}

The concepts from information theory and Luhmann's social theory which we use to model expectation-driven interaction are:
\begin{enumerate}
	\item Information and meaning structure,
	\item Interaction, and
	\item Double contingency and expectation-expectation.
\end{enumerate}
These concepts will be integrated in a model which explores interaction strategies using different types of interconnected memories. The interconnectedness of the memories 
is a precondition for obtaining agents ``capable of acting.'' 

\subsection{Information and meaning structure.}\label{sec:infomeaningstructure}

The concept of information used in this paper originates with the mathematical theory of communication of Shannon and Weaver \citeyearpar{Wea:1949}, now known as information theory.  First, information is defined as being produced when one message is chosen from a set of messages \citep{Sha:1948}. The minimum requirement of information production is therefore a set of two messages and a ``choice'' (usually, the number of messages is higher, but finite). 
Second, information sources are characterized by entropy:
\begin{quote}
That information be measured by entropy is, after all, natural when we remember that information, in communication theory, is associated with the amount of freedom of choice we have in constructing messages.
Thus for a communication source one can say, just as he would also say it of a thermodynamic ensemble, `This situation is highly organized, it is not characterized by a large degree of randomness or of choice---that is to say, the information (or the entropy) is low.'
\hfill\citep[pg.~13]{Wea:1949}
\end{quote}

This means that the greater the freedom of choice and consequently the greater the amount of information, the greater is the uncertainty that a particular message will be selected. Thus, greater freedom of choice, greater uncertainty and greater information go hand in hand. It is important to note that this uncertainty-based concept of information must not be confused with meaning. Although messages usually do have meaning, this is irrelevant to the mathematical formulation of information on the basis of uncertainty and entropy. Two messages, one heavily loaded with meaning and the other of pure nonsense, can have the same amount of information.

The theory of social systems refers to Shannon and Weaver and uses as a starting point the same definition: information is a selection from a repertoire of possibilities \citep[pg. 140]{Luh:1995}. But contrary to information theory, meaning is addressed prominently. It is conceptualized as the durable complement of the transient information. Meaning is a complex referential structure which is used by psychic systems (\ie, information sources) to organize the process of information selection.
\begin{quote}
By information we mean an event that selects system states. This is possible only for structures that delimit and presort possibilities.
Information presupposes structure, yet is not itself a structure, but rather an event that actualizes the use of structures\ldots. Time itself, in other words, demands that meaning and information must be distinguished, although all meaning reproduction occurs via information (and to this extent can be called information processing), and all information has meaning\ldots. a history of meaning has already consolidated structures that we treat as self-evident today.
\hfill\citep[pg.~67]{Luh:1995}
\end{quote}
According to social systems theory, the historically evolved general meaning structure is  represented on the micro level (\ie, psychic systems, agent level) in the form of the personal life-world \citep[pg.~70]{Luh:1995}. A personal meaning-world represents a structurally pre-selected  repertoire of possible references. Although the repertoire of possibilities is limited, selection  is necessary to produce information. Meaning structure provides a mental map for selection  but does not replace it. 
All together, the one is not possible without the other---information production presupposes meaning structure and the actualization of meaning is done by information production.

The present model considers both information and personal meaning structure in a simplified manner. Information is treated in the above described mathematical form and meaning structures in the form of a second order approach (pairwise combination). This will be developed in \secref{sec:modeldef}, below.

\subsection{Interaction and interaction sequence.}\label{sec:msgexchangecomm}

Communicative interaction depends on the availability of information producing agents. Individuals or other agents are therefore often the elemental unit of sociological communication analysis. Contrary to that, communication units are the central element of analysis in social system theory and agents are conceptualized according to their contribution to communication units. Each communication unit consists of three selections \citep{Luh:1995}: 
\begin{enumerate}
	\item Information is produced by selection from a structured repertoire of possibilities by an agent (alter).
	\item The agent (alter) must choose a behavior that expresses the produced information (act of utterance).
	\item Another agent (ego) must understand what it means to accept alter's selection.
\end{enumerate}

Information production includes therefore  that alter not only views itself as part of  a general meaning world but assumes also that the same is true for ego. In other words, alter and ego act on the assumption that each life-world is part of the same general meaning world in which information can be true or false, can be uttered, and can be understood. This assumption of shared meaning structures allows alter to generate an expectation of success, \ie, the expectation that the selection will be accepted by ego.
\begin{quote}
By `success' I mean that the recipient of the communication accepts the selective 
content of the communication (the information) as a premise of his own behavior, thus 
joining further selections to the primary selection and reinforcing its selectivity in the 
process.
\hfill\citep[pg.~88]{Luh:1990}
\end{quote}

An interaction system is constituted by one communication unit following another. The closure of an interaction system is given by two constituents \citep[pg.~5ff]{Luh:1990}: 
participants must be present and communication units can only be understood in the
context of the system. Consequently, 
at least a minimum extent of understanding is necessary for the generation of an 
interaction system. If one communication unit directly follows another, it is a positive test that the preceding communication unit was understood sufficiently. Every communication unit is therefore recursively secured in possibilities of understanding and the control of understanding in the connective context. 

In the present model, the term ``message'' is used to  describe the three parts of the communication unit. From this follows that the model is at its core a message exchange model (ME model) and starts with a list of messages. The three-part unity is now: 
\begin{enumerate}
	\item Alter selects a message.
	\item Once the selection process is finalized, the message is sent. 
	\item Ego receives the sent message. 
\end{enumerate}
Within the present version of the ME model, agents do not have the option to reject a message or to refuse to answer. Their ``freedom'' is incorporated in the process of message selection. Agents can be distinguished by the number of messages they are able to use and by their selection strategies. The ongoing exchange of messages creates an interaction sequence 
with a limited---but possibly high---number of steps.

Overall, the term ``interaction'' is considerably simplified in comparison to the full Luhmannian notion. Additionally, the notion of interaction system is not 
used in the present version of the model, because ``understanding'' is not considered.

\subsection{Double contingency and expectation-expectation.} 
\label{sec:expectationstructure}

In social system theory, the dynamics of an interaction sequence are explained by contingency and expectation\footnote{Expectations in social systems theory are different from expectations in probability 
or statistics; the meaning is closer to the statistical notion of conditional expectation. In this work, we use the term in the social systems sense.}. First, contingency arises because agents 
are complex systems that are ``black boxes'' for each other. An agent will never know exactly what the other will do next. The reciprocity of social situations (two agents: alter and ego) results in  double contingency: alter contingency and ego contingency.   

Second, the black-box problem is solved by expectations. Agents create expectations about  the future actions of the other agents to adjust their own actions. For example, two business  partners meeting antemeridian expect from each other that the adequate greeting is good  morning (and not good evening, \etc). This is a situation with high expectational security. If  the two had a conflict last time they met, the expectational security may decrease (will the  other answer my greeting?). The example illustrates that expectations are condensed  forms of meaning structures \citep[pp.~96, 402]{Luh:1995}. Based on the societally given and biographically determined structures of expectations, alter and ego have situation-specific reciprocal expectations.

\begin{quote}
In social systems, expectations are the temporal form in which structures develop. But as structures of social systems expectations acquire social relevance and thus suitability only if, on their part, they can be anticipated. Only in this way can situations with double contingency be ordered. Expectations must become reflexive: it must be able to relate to itself, not only in the sense of a diffuse accompanying consciousness but so that it knows it is anticipated as anticipating. This is how expectation can order a social field that includes more than one participant. Ego must be able to anticipate what alter anticipates of him to make his own anticipations and behavior agree with alter's anticipation.
\hfill\citep[pg.~303]{Luh:1995}
\end{quote}

The duplication of contingency causes the duplication of expectation: 
\begin{itemize}
\item Alter has expectations vis-\`a-vis ego. 

\item Ego knows that alter expects something from him/her, but can never know what exactly the expectation is.

\item Therefore, ego builds expectations about ``alter's expectations of ego,'' that is, expectation-expectation. 
\end{itemize}

In complex societies, expectational security is difficult. Therefore, it is necessary to have stabilizing factors. One way to establish expectations that are relatively stabilized over time is through relation to something which is not itself an event but has duration. Examples of well-known expectational nexuses in sociology are persons and roles. Persons are societally constituted for the sake of ordering of behavioral expectations. Roles serve---compared with persons---as a more abstract perspective for the identification of expectations. 

Within the ME model in its present form, the agents are the relevant expectational nexuses. 

\subsection{Memories and memory development.} \label{sec:learning}

The agents are equipped with three different types of memories: initially, their behavior
tends towards 
a general long-term memory which represents a given scope of interaction. That is, 
they are able to use a certain number of messages in interaction sequences. The ability to 
use messages is determined by the above mentioned pairwise combination. The model agents
start with a randomly or otherwise specified stimulus-response matrix (response 
disposition, see \secref{sec:memories}). 

Since memory development depends on expectation driven interaction, 
agents are equipped with specific memories allowing them to construct expectations. 
This is done by the introduction of specific ego- and alter-memories \citep[following][]{DitKroBan:2003}. The ego memory represents an agent's expectation towards itself and the alter 
memory an agent's expectation-expectation (\ie, expectations of \agent{B} about the 
expectations of \agent{A} vis-\`a-vis \agent{B}). The model starts with pristine ego- and alter-
memories which will be updated with every interaction. The links between 
the long-term memory and ego- and alter-memory will be explained in \secref{sec:modeldef}.

\subsection{Interaction strategies.} \label{sec:strategies}

Within the presented model, interaction strategies are selection strategies. All three memories are used in each information-production activity (\ie, selection from the pairwise structured repertoire of possibilities in the form of stimulus-response matrices). The way the memories are used defines the interaction strategy. Agents can be specified to follow various strategies, \eg, ``conservative'' agents maximize towards predictability, ``innovative'' agents towards unpredictability.

Interaction strategies are formalized in \secref{sec:transmatrix}. The strategies result from  balancing out conflicting behavioral intentions for agent interactions.  Specifically: 
(1) ego-oriented behavior, maintaining individual behavior patterns \emph{disregarding} the interaction partner; 
(2) interaction-oriented behavior, providing  stable responses \emph{for} the interaction partner through consistent and comprehensible behavior;  
(3) uncertainty-oriented behavior,  seeking or avoiding unexpected responses \emph{from} the interaction partner; and
(4) risk-oriented behavior, handling  novel or erroneous responses. 
We investigate the effect of various strategies using analytical (\secref{sec:mathsummary}) and computational (\secref{sec:simresults}) methods. 

In this work, we focus on how strategies affect interaction; we will not present rules for choosing strategies. However, it is clear that the choice of a strategy is a way for agents to cope with the problem of appropriate behavior in a situational context. A desirable---and challenging---extension of the presented model would be to develop rules for choosing and evaluating strategies.

\section{Model definition.} \label{sec:modeldef}

\subsection{Messages.} \label{sec:messages}

In the model,
the fundamental elements of agent interactions are
messages. 
The number of messages
is assumed to be finite. 
We do not concern ourselves here with giving a real-world interpretation to the messages, and
instead simply number them from 1 to  the number of messages \(\nummessages\). 

Messages are transmitted through a multi-step process. A message is first selected,
then sent and simultaneously  received by the interaction partner. We do not consider  the medium
in which the message is sent. Following \citet{DitKroBan:2003}, we assume that messages are
distinct from one another. This assumption provides a 
significant simplification over, \eg, natural languages, ignoring issues such as ambiguity or transmission errors. 
 
It can be convenient to recognize that 
messages transmitted by an agent are formally equivalent to states 
of the agent. A construction by 
\citet{DitKroBan:2003} illustrates this in a
readily accessible fashion: each agent has a number of signs available,
and communicates a message by holding up the appropriate sign. 
Thus, selecting a new message to transmit 
is equivalent to a state transition---the agent holding up a new sign.

\subsection{Interaction sequences.} \label{sec:commseq}

Agents take turns sending messages to each other. In doing so, they
record the foregoing sequence of messages, and must select the
next message in the sequence. 

We anticipate that the message selected will depend most strongly on
the most recent messages in the sequence, while the exact ordering of
earlier messages is of lesser or negligible importance. This suggests that a 
reasonable approximation of the selection process is to consider only a 
finite number of preceding messages, rather than the complete sequence. 
Thus, the agents should select the message based on the statistics of 
subsequences of various lengths. Earlier portions of the message sequence
are used to define the agents and the rules by which they select messages (see
\secref{sec:agents}).

The length of the subsequences in 
such approximations define the order of the
approximation. A zeroth-order 
approximation considers only message identity, ignoring how the messages
actually appear in sequences. A first-order approximation 
considers message sequences of length one, showing the relative frequency with 
which the messages appear in  sequences. Second- and 
higher-order approximations consider runs of messages, showing 
correlations between messages in the sequences.

In this paper, we focus exclusively on a second-order approximation,
dealing with pairs of messages.
A second-order approximation 
is the simplest case in which relationships between messages become
important, and can be usefully interpreted as dealing with
stimuli and responses. However, it is vitally important to recognize that 
each message is simultaneously used as both stimulus and response. 

In this work, we adopt the point of view that all agent interactions are expressed and interpreted in
terms of message sequences. Consequently, we assess interactions in terms of message sequences as well. An example of how to do this is to   identify a specific agent's behavior as authoritative, effectively defining a utility function; the student-teacher situation 
described in \secref{sec:learnrates} is analyzed in this way.
Interaction of a single agent with its environment can be treated in the same
fashion by describing the environment itself as a second agent and the interaction as messages. 

\subsection{Agents.} \label{sec:agents}

In this section, we formally define the agents. Essentially, an agent 
maintains a number of memories that describe statistics of message
sequences in which it has taken part, and has a rule for using those 
memories to select a response message to a stimulus message.  The 
agents are a generalization of the agents used by \citet{DitKroBan:2003}; see
\secref{sec:transmatrix}.

Although we focus in this work on interactions between pairs of agents, the agent definitions are posed in a manner that can be generalized to communities of agents. See \secref{sec:multiagent} for discussion of how the model could be extended to handle groups of agents.

\subsubsection{Memories.} \label{sec:memories}

Agent memories fall broadly
into two classes: (1)  memories of the details of 
particular interaction sequences between
one agent (alter) and another agent (ego)
and (2) long-term memories reflecting the general 
disposition of a single agent. All of the memories are represented
as probability distributions over stimuli and responses. 

We use three different forms of probabilities in the definition and application of 
the agent memories. For example, consider the ego memory (defined fully below).  
We denote by \(\egomemjoint{t}{A}{B}{r}{s}\) the joint probability that \agent{A} maintains 
at time~\(t\), estimating the likelihood that \agent{B} will transmit a message~\(s\) to which \agent{A} will respond with a message~\(r\). Second, we use the conditional 
probability~\(\egomemcond{t}{A}{B}{r}{s}\), showing the likelihood of \agent{A} responding with message~\(r\) given that \agent{B} has already transmitted message~\(s\). Finally, we 
use the marginal probability~\(\egomemmarg{t}{A}{r}\), showing the likelihood of \agent{A} responding with message~\(r\) regardless of what message \agent{B} sends. These three probabilities are related by 
\begin{eqnarray}
	\egomemmarg{t}{A}{r} & = & \sum_{s}\egomemjoint{t}{A}{B}{r}{s} \\
	\egomemcond{t}{A}{B}{r}{s} & = & 
		\frac{\egomemjoint{t}{A}{B}{r}{s}}{\sum_{r}\egomemjoint{t}{A}{B}{r}{s}}
	\mathperiod
\end{eqnarray}
We use similar notation for all the memories (and other probabilities) in this work.

The memories of messages exchanged are patterned after
those used by \citet{DitKroBan:2003}. 
They take two complementary forms, the ego memory and
the alter memory, reflecting the roles of the agents as
both sender and receiver.

The ego memory \(\egomemjoint{t}{A}{B}{r}{s}\) is a 
time-dependent memory that 
\agent{A} maintains about interactions with \agent{B}, where 
\agent{B} provides the stimulus \(s\) to which \agent{A} gives 
response \(r\). During the course of interaction
with \agent{B}, the corresponding ego memory
is continuously updated 
based on the notion that
the ego memory derives from the relative frequency of stimulus-response
pairs. In general, an agent will have a distinct ego memory for each
of the other agents.

Before \agentsxandy{A}{B} have interacted, \agent{A}'s
ego memory
\(\egomemjoint{t}{A}{B}{r}{s}\) (as well as \agent{B}'s 
counterpart \(\egomemjoint{t}{B}{A}{r}{s}\))
is undefined. 
As \agentsxandy{A}{B} interact through time \(t\), they together 
produce a message sequence 
\(\messagesequence{1}{2}{t-1}{t}\).
Assuming that \agent{B} has sent the first message, one view
of the sequence is that \agent{B} has provided a sequence of stimuli
\(\messagesequence{1}{3}{t-3}{t-1}\)
to which \agent{A} has provided a corresponding set of responses 
\(\messagesequence{2}{4}{t-2}{t}\). 
If \agent{A} began the communication,
the first message could be dropped from the communication sequence; similarly,
if \agent{B} provided the final message, this ``unanswered stimulus'' could be 
dropped from the communication sequence.

With this view of stimuli and responses, the ego 
memory for \agent{A} is defined by
\begin{equation}
    \egomemjoint{t}{A}{B}{r}{s} = \frac{2}{t} \sum_{i=1, 3, 5, \ldots}^{t-1} 
                \krondelta{s}{m_{i}}\krondelta{r}{m_{i+1}}
    \mathcomma
    \label{eq:egomemdef}
\end{equation}
for all \(r\) and \(s\). 
In \eqnref{eq:egomemdef}, we assume that the memory has an infinite 
capacity, able to exactly treat any number of stimulus-response pairs.
A natural and desirable modification is to consider memories with 
a finite capacity. 

Limited memory could be incorporated in a variety of ways. The approach we take 
is to introduce a ``forgetting'' parameter \(\egoforgetparam{A}\), with a value from the interval \([0, 1]\), such that the 
ego memory is calculated as
\begin{equation}
    \egomemjoint{t}{A}{B}{r}{s} = 2\frac{1-\egoforgetparam{A}}{1-(\egoforgetparam{A})^{t}} 
    				\sum_{i=1,3,5,\ldots}^{t} (\egoforgetparam{A})^{t-i-1} 
				\krondelta{r}{m_{i+1}}\krondelta{s}{m_{i}}
    \label{eq:egomemleakydef}
\end{equation}
for all \(r\) and \(s\). The memory of a particular message transmission decays exponentially.

The alter memory \(\altmemjoint{t}{A}{B}{r}{s}\) is analogous to the ego 
memory, but with the roles of sender and receiver reversed.
Thus, \(\altmemjoint{t}{A}{B}{r}{s}\) is the memory that 
\agent{A} maintains about message exchanges with \agent{B}, where 
\agent{A} provides the stimulus \(s\) to which \agent{B} gives 
response \(r\). The procedure for calculating the alter memory 
directly parallels that for the ego memory, except that the messages
sent by \agent{A} are now identified as the stimuli and the messages
sent by \agent{B} are now identified as the responses. The calculation
otherwise proceeds as before, including making use of a forgetting parameter \(\altforgetparam{A}\) for the alter memory.

A useful symmetry often exists between the ego and alter memories of the agents. Using the alter memory \(\altmemjoint{t}{A}{B}{r}{s}\), \agent{A}  tracks the
responses of \agent{B} to stimuli from \agent{A}. This is exactly what \agent{B} tracks in its ego memory \(\egomemjoint{t}{B}{A}{r}{s}\). Thus, the memories of the two agents are related by 
\begin{equation}
	\altmemjoint{t}{A}{B}{r}{s} = \egomemjoint{t}{B}{A}{r}{s}
	\mathperiod
	\label{eq:memsymmetry}
\end{equation}
However, \eqnref{eq:memsymmetry} holds only if the agent memories are both infinite or \(\altforgetparam{A}= \egoforgetparam{B}\). The corresponding relation \(\altmemjoint{t}{B}{A}{r}{s} = \egomemjoint{t}{A}{B}{r}{s}\) holds as well when the memories are both infinite or \(\egoforgetparam{A} = \altforgetparam{B}\).

Besides the ego and alter memories, an agent has another 
memory called the response disposition. The response 
disposition \(\responsedispjointnoargs{A}\) of \agent{A} 
is, like the ego and alter memories, represented as a joint 
distribution, but there are many marked differences. Significantly,
the response disposition is not associated with another particular 
agent, but instead shows what the agent brings to all interactions. In particular, the response disposition can 
act like a general data bank applicable
to  any other agent. Further, 
the response disposition changes more slowly. Instead, the update occurs 
after the interaction sequences. In this work, we hold the response dispositions fixed, so we defer discussion of a possible update rule for the 
response disposition to \secref{sec:assess}.

\subsubsection{Transition probability.} \label{sec:transmatrix}

The memories of an agent are 
combined to produce a transition probability, 
which is, in turn, used to randomly select messages.
The form of the transition probability is a generalization of that used
by \citet{DitKroBan:2003}.  
Differences arise from the differing goals of the two models:
\citeauthor{DitKroBan:2003} consider the formation of social order, while
we are interested in strategies of interaction.

The transition 
probability is constructed so as to deal with several distinct  issues. 
Because of this, we do not develop the transition
probability as an optimal selection rule according to one criterion, but
rather as a compromise between several conflicting factors
arising in a situation of double contingency. We consider three
such  factors: 
(1) agent preference,
(2) satisfaction of expectations, and
(3) treatment of individual uncertainty. 

The first factor  that we consider is agent preference. 
An agent tries to select messages that are  
correct for the situation, based on its prior experiences. The response disposition matches well with
this notion, since, as discussed above, it is a relatively stable memory of the general 
agent behavior.   

The second factor that  we consider is that an agent tries to satisfy the expectations that its interaction 
partner has about the agent itself. This is the 
expectation-expectation.
Consider interactions between \agentsxandy{A}{B}.
While selecting which message to send, \agent{A} estimates what 
message \agent{B} would expect to receive from \agent{A}. 
In terms of the memories of \agent{A}, this is expressed as
the ego memory \(\egomemjoint{t}{A}{B}{r}{s}\). Thus, what responses
\agent{A} has given to \agent{B} in the past will be favored in the 
future, so that there is a continual pressure for an agent to be
consistent in its responses.

The third factor  we consider is that  an agent takes into account the level of certainty of the expected response from its interaction partner. By favoring messages that lead to unpredictable responses, the agent can pursue ``interesting,''  interactions of greater complexity. Alternatively, the agent may favor messages that lead to predictable responses, pursuing simpler (and more comprehensible) interactions.

A natural
approach is for \agent{A} to calculate, based on previous responses of \agent{B}, 
the uncertainty for each
possible message \(r\) that it could send.
This corresponds to calculating the entropy based on the 
conditional alter memory \(\altmemcond{t}{A}{B}{\rprime}{r}\), yielding
\begin{equation}
    \respentropy{t}{A}{B}{r} = \entropybasen{\altmemcond{t}{A}{B}{\rprime}{r}}
    						{\rprime}{\nummessages}
    \mathperiod
    \label{eq:uncertainty}
\end{equation}
The base of the logarithm is traditionally taken as 2, measuring the entropy in bits, but in this work we take the base to be \(\nummessages\), the number of different messages. With this choice, the entropy take on values from the interval \([0, 1]\), regardless of the number of messages.

The uncertainty \(\respentropy{t}{A}{B}{r}\) is closely related to the 
expectation-certainty from \citet{DitKroBan:2003}.
Note that all possible messages \(r\) are considered, so that \agent{A}
may send a message to \agent{B} that is highly unlikely  
based on a just-received stimulus. Thus, we expect that 
resolving uncertainty and satisfying expectations will come into conflict.

The transition probability is assembled from the foregoing components 
into a conditional probability distribution with form
\begin{equation}
    \selectprobcond{t}{A}{B}{r}{s} \propto 
                               \knowledgeparam{A} \responsedispcond{A}{r}{s}
    			 + \expectationparam{A} \egomemcond{t}{A}{B}{r}{s}
                                + \uncertaintyparam{A} \respentropy{t}{A}{B}{r}
                                + \randomparam{A} \frac{1}{\nummessages}
    \mathperiod
    \label{eq:transitionprobability}
\end{equation}
The parameters \knowledgeparam{A}, \expectationparam{A},  and \uncertaintyparam{A}  
reflect the relative importance of the three factors discussed above, while 
\randomparam{A} is an offset that provides a randomizing element. The
randomizing element provides a mechanism  for, \eg, introducing novel 
messages or transitions.  It also plays an important role on mathematical
grounds (see the appendix).  

Note that we allow the parameter \(\uncertaintyparam{A}\)  to be negative.
A positive \(\uncertaintyparam{A}\)  corresponds to \agent{A} favoring messages that lead to unpredictable responses,
while a negative \(\uncertaintyparam{A}\) corresponds to \agent{A} ``inhibiting'' messages that lead to unpredictable response,
hence favoring messages leading to predictable responses.
When \(\uncertaintyparam{A}\) is negative, \eqnref{eq:transitionprobability} could become negative.  Since 
\(\selectprobcond{t}{A}{B}{r}{s}\) is a probability, we deal with this negative-value case
by setting it to zero whenever  \eqnref{eq:transitionprobability} gives a negative value.

Clearly, one may expect different results in the outcome of interactions when   \(\uncertaintyparam{A}\)
is changed from a positive to a negative value. This is illustrated mathematically in the appendix and numerically
in \secref{sec:simresults}. We do, in fact, observe a sharp transition in the entropy of the alter and ego memories of the agent (measuring the average uncertainty in their response)  near \(\uncertaintyparam{A} = 0\).
Note however that the transition depends also on the other parameters and does not necessarily occur exactly at
\(\uncertaintyparam{A}=0\). This is discussed in the appendix.   
The transition is  reminiscent of what \citeauthor{DitKroBan:2003} 
call the appearance of social order, though our model
is different, especially because of the response disposition which plays a crucial role in the transition.

Appropriate choice of the coefficients \(\knowledgeparam{A} \),
\(\expectationparam{A} \), \(\uncertaintyparam{A}\), and \(\randomparam{A}\) allows the transition probabilities used by \citeauthor{DitKroBan:2003} to be constructed as a special case of \eqnref{eq:transitionprobability}. In particular, their parameters \(\alpha\), \(c_{f}\), and \(N\) are related to the ones used here by
\begin{eqnarray}
	\knowledgeparam{A} & = & 0 \\
	\expectationparam{A} & = & 1-\alpha\\
	\uncertaintyparam{A} & = & -\alpha\\
	\randomparam{A} & = & \nummessages\left(\alpha + \frac{c_{f}}{N} \right )
	\mathperiod
	\label{eq:dkbconnection}
\end{eqnarray}

Note finally that the transition probability given in \eqnref{eq:transitionprobability} is not appropriate for selecting the initial message of an interaction sequence. 
However, the response disposition provides a useful mechanism for selection of the initial message. We calculate a marginal probability for the response \(r\) from the  response disposition \(\responsedispjoint{A}{r}{s}\), using
\begin{equation}
	\responsedispmarg{A}{r} = \sum_{s}\responsedispjoint{A}{r}{s}
	\mathperiod
	\label{eq:initmessageprobability}
\end{equation}
With the marginal probability of initial messages \(\responsedispmarg{A}{r}\) and the conditional probability for responses \(\selectprobcond{t}{A}{B}{r}{s}\), stochastic simulations of the model system are relatively straightforward to implement programmatically.

\section{Results.}\label{sec:results}

\subsection{Summary of mathematical results.} \label{sec:mathsummary}

In this section we briefly summarize the mathematical results obtained
in the paper. A detailed description of the mathematical setting is given in the appendix.
Supporting numerical results are presented in \secref{sec:simresults}.

The model evolution is a stochastic process where the probability for an agent to select a message  is given by  a transition matrix, as in a Markov process.
However, the transition matrix is history-dependent and the corresponding process is not Markovian. 
The general mathematical framework for this type of process
is called ``chains with complete connections.'' \citep{Mai:2003}
A brief overview is given in the appendix.

On mathematical grounds, a precise description of the pairwise interaction between agents requires answering at least the following questions:
\begin{enumerate}
	\item Does the evolution asymptotically lead to a stationary regime ?
	\item What is the convergence rate and what is the nature of the transients?
	\item How can we quantitatively characterize the asymptotic regime?
\end{enumerate}
We address these briefly below. Many other questions are, of course, possible.

\emph{Does the evolution asymptotically lead to a stationary regime?}
In the model, convergence to a stationary state means that, after having exchanged
sufficiently many symbols, the ``perceptions''that each agent has of the other's behavior (alter memory)
and of its own behavior (ego memory) have stabilized and
do not evolve any further. 
The probability that \agent{A} provides a given symbol as a response to some stimulus
from \agent{B} is not changed by further exchanges.     

Mathematically, the convergence to a stationary state  is not straightforward.
One cannot \apriori\ exclude non-convergent regimes,
such as cycles, quasi-periodic behavior, chaotic or intermittent evolution, \etc. 
A cyclic evolution, for example, could correspond to a cyclothymic agent whose reactions vary periodically in time.
More complex regimes would essentially correspond to agents whose reactions evolves in an unpredictable
way. Though such behaviors are certainly interesting from a psychological point of view, they are not
desirable in a model intended to describe knowledge diffusion in a network of
agents assumed to be relatively ``stable.'' It is therefore desirable to identify conditions
that ensure the convergence to a stationary state. These conditions are discussed in the appendix.
We also establish explicit equations for the memories in the stationary state.

\emph{What is the convergence rate and what is the nature of the transients?} The answer to this question is as important
as the previous one. Indeed, the investigated asymptotic behavior is based on the limit where interaction sequences are infinitely long,
while in real situations they are obviously finite. However, if the characteristic time necessary for an agent
to stabilize its behavior is significantly shorter than the typical duration of an interaction,
one may consider that the asymptotic regime is reached and the corresponding results essentially apply.
In the opposite case, the behavior is not stabilized and important changes in the stimulus-response
probabilities may still occur. This transient regime is more difficult to handle and we
have focused on the first case. The characteristic time for the transients depends on the parameters \(\knowledgeparamtrunc\) through  \(\randomparamtrunc\).

\emph{How can we quantitatively characterize the asymptotic regime?}
Addressing this question basically requires identifying a set of relevant observables---that is,
a set of functions assigning to the current state of the two-agent system a number that corresponds
to a rating of the interaction. We have obtained some mathematical and numerical results
about the variations of the observables when the coefficients 
in \eqnref{eq:transitionprobability} are changed. The most salient
feature is the existence of a sharp transition when \(\uncertaintyparamtrunc\) changes its sign. Simulations demonstrating the transition are presented in \secref{sec:statbehavior}.

We have chosen several observables.  The \emph{entropy} of a memory  measures the uncertainty or unpredictability of the stimulus-response relationship. A memory with high entropy is representative of complex (and, presumably, hard to understand) interaction sequences. We inspect the entropy of both the ego and alter memories.
The \emph{distance} or overlap between two memories
measures how close are their representations of the stimulus-response relationship. Particularly interesting is the distance between the response disposition of an \agent{A} and the alter memory another \agent{B} maintains about \agent{A}, which shows the discrepancy between the actual behavior of \agent{A} and  \agent{B}'s expectation of that behavior. Note that
this representation is necessarily biased since \agent{B} selects the stimuli it uses to acquire its knowledge about \agent{A}. 
This induces a bias (see appendix for a mathematical formulation of this).

To recap, in the appendix we provide a mathematical setting where we discuss the convergence of the dynamics to a
state where memories are stabilized.
We obtain explicit equations for the asymptotic state and discuss
how the solutions depend on the model parameters. In particular, we show that a drastic change occurs when
\(\uncertaintyparamtrunc\) is near zero. We assign to the asymptotic state a set of observables
measuring the quality of the interaction according to different criteria.

\subsection{Simulation results.} \label{sec:simresults}\label{NumSim}

\subsubsection{Types of agents.} \label{sec:agenttypes}

A broad variety of agents can be realized from the definitions in \secref{sec:agents}.
The value used for the response disposition reflects an individualized aspect of an agent's behavior, while the various other parameters allow a wide range of strategies for interaction. 
In \eqnref{EABSolFinal}, we give an analytic expression for the memories in the 
asymptotic regime.
However, an extensive study of the solutions of this equation, depending on six
independent parameters,
is beyond the scope of this paper. Instead,  we focus in this section on a few situations  with different types of agents, as characterized by their response dispositions.

\begin{figure}[htbp]
	\centering
	\includegraphics[width=\textwidth]{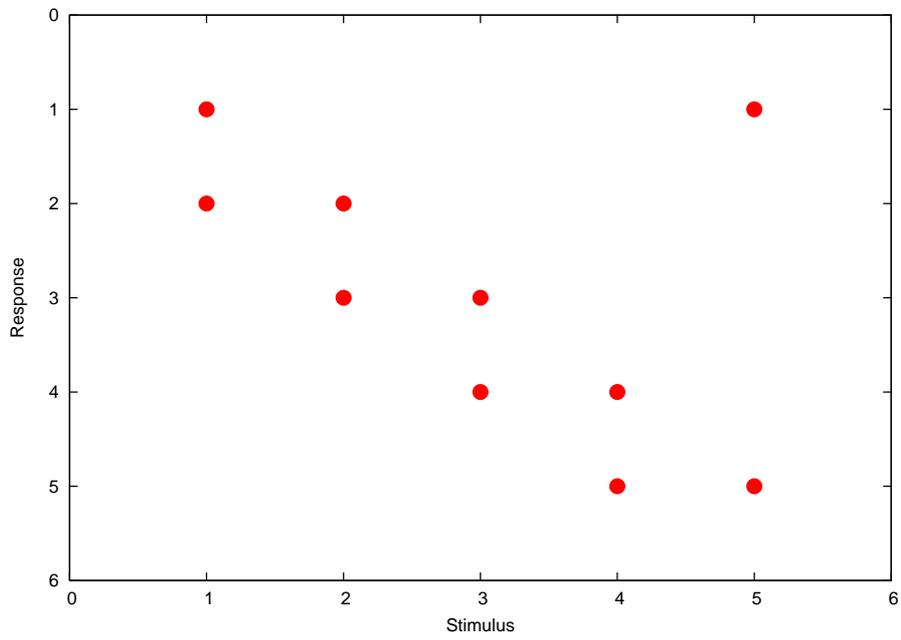}
	\caption{Banded response disposition. The size
	of the circle is proportional to the value of the corresponding entry.}
	\label{fig:bandeddisp}
\end{figure}

One type of agent has a random response disposition matrix,  created by selecting each element
uniformly from the interval \([0,1]\) and normalizing the total probability. With random response dispositions of this sort, the contribution to the agent behavior is complex but unique to each agent.
A second type of agent has its behavior restricted to a subset of the possibilities, never producing certain messages. A response disposition for this type of agent, when presented as a matrix, can be expressed as a block structure, such as in \figref{FigDisp}. A third and final type of agent response disposition that we consider has a banded structure, as 
in \figref{fig:bandeddisp}, allowing the agent to transmit any message, but with restricted transitions between them.

The parameters \knowledgeparam{A}, \expectationparam{A}, \uncertaintyparam{A}, and~\randomparam{A} also play important roles in the behavior of an \agent{A}. We impose a constraint with the form
\begin{equation}
	\knowledgeparam{A} + \expectationparam{A} + 
	\uncertaintyparam{A} + \randomparam{A} = 1
	\mathperiod
	\label{eq:paramconstraint}
\end{equation}
The constraint is in no way fundamental, but is useful for graphical presentation of simulation results. 

\subsubsection{Complementary knowledge.} \label{sec:complementaryknowledge}

\begin{figure}[thbp]
	\centering
	\subfigure[\agent{A}]{\includegraphics[width=6cm]{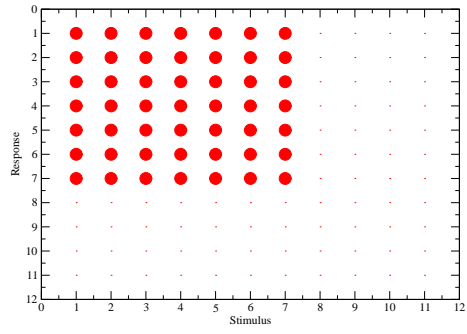}
			  	\label{fig:alicedisp}}
	\subfigure[\agent{B}]{\includegraphics[width=6cm]{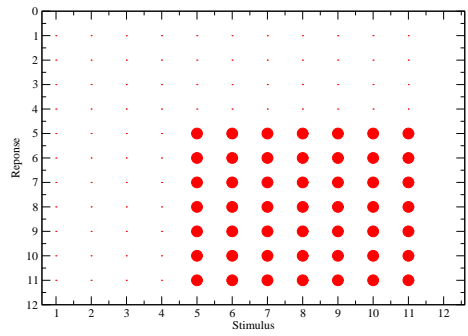}
				\label{fig:bobdisp}}
	\caption{Block-structured response dispositions. The size
	of the circle is proportional to the value of the corresponding entry.}
	\label{FigDisp}
\end{figure}

As an initial simulation, we consider interactions between two
agents with block-structured response dispositions, as shown in \figref{FigDisp}.  
The number of messages is  \(\nummessages=11\). The elements of the probability matrices
 are shown as circles whose sizes are proportional to the value of the matrix elements,
 with the sum of the values equal to one.
 
 The situation
shown in \figref{FigDisp} corresponds to a case of ``complementary
knowledge,'' where the agent behaviors are associated with distinct subdomains of the larger context. Each response disposition matrix has a \(7 \times 7\) block where the elements have 
uniform values of \(1/49\), while all other elements are zero. The blocks for the two agents have 
a non-zero overlap.  The subdomain for \agent{A} deals with messages 1 through 7,
while that for  \agent{B} deals with messages 5 through 11.
Thus, if \agent{B} provides a stimulus from \(\set{5, 6, 7}\), \agent{A} responds with a message
from \(\set{1, 2, \ldots, 7}\) with equiprobability. The coefficients \(\randomparam{A}\) and \(\randomparam{B}\)
are both set to \(0.01\). These terms are useful to handle the response of the agent when
the stimulus is ``unknown'' (\ie, it is not in the appropriate block). In such a case, the agent responds to any stimulus message with equal likelihood.


Interaction between the two agents consists of a large number of short interaction sequences. We study the evolution of the alter and ego memories of each agent as the 
number of completed interaction sequences increases. 
Each interaction sequence consists of the exchange of 11 messages between the two agents, with the agents taking part in 128000 successive interaction
sequences.  For each interaction sequence,  one of the two agents is chosen to select the initial message at random, independently of how previous sequences were begun  
(randomly choosing this ``starting agent'' is intended to avoid pathological behavior
leading to non-generic sequences).

We first checked the convergence of \(\egomemmarg{t}{A}{r}\) to the first mode of \(\selectprobcondnoargs{t}{A}{B}\selectprobcondnoargs{t}{B}{A}\)
(see \appendref{sec:stationarystateequations}).
We measure the distance between these two vectors at the end of the simulation. The distance
tends to zero when the number of interaction sequences \(T\) increases, but it decreases slowly, like \(1/T\)
(as expected from the central limit theorem). An example is presented
in \figref{FigConv} for \(\knowledgeparam{A}=\knowledgeparam{B}=0.99\) and
\(\expectationparam{A} = \expectationparam{B} = \uncertaintyparam{A} = 
\uncertaintyparam{B} = 0\).


We next investigated four extremal cases 
\((\knowledgeparam{X}=0.99; \expectationparam{X}=0; 
   \uncertaintyparam{X}=0; \randomparam{X}=0.01)\),
\((\knowledgeparam{X}=0; \expectationparam{X}=0.99; 
   \uncertaintyparam{X}=0; \randomparam{X}=0.01)\), 
\((\knowledgeparam{X}=0; \expectationparam{X}=0; 
   \uncertaintyparam{X}=0.99; \randomparam{X}=0.01)\),
and
\((\knowledgeparam{X}=0.99; \expectationparam{X}=0.99; 
   \uncertaintyparam{X}=-0.99; \randomparam{X}=0.01)\). 
For each case, we first examined the convergence
to the asymptotic regime. We show the evolution of the joint entropy in 
\figref{FjointS}. This allows us to estimate the time needed to reach the 
asymptotic regime.

\begin{figure}[thbp]
	\centering
	\includegraphics[width=10cm,clip=false]{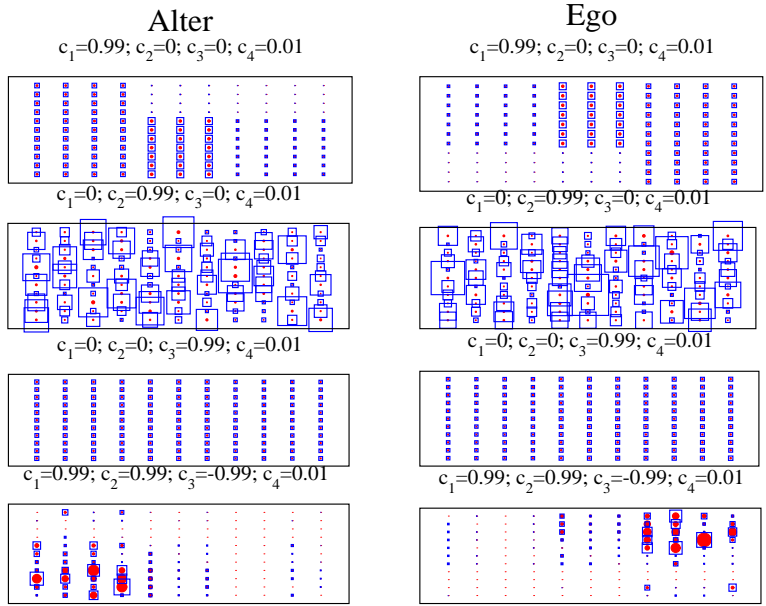}
	\caption{Average asymptotic memories for \agent{A}
	after $128000$  steps, 
	with infinite memories.
	The matrices shown here are calculated by averaging over $10$  initial conditions.
	The sizes of the red circles are proportional to the corresponding matrix elements, 
	while the sizes of the blue
	squares are proportional to the mean square deviations.}
	\label{FMatrAs2}  
\end{figure}

Next, we plotted the asymptotic values of the ego and alter memory of \agent{A}, averaged over several samples. They  are shown in \figref{FMatrAs2} for \(\altforgetparamabbrev = \egoforgetparamabbrev = 1\).  In \figref{FMatrAs4},
we show the values with  
a finite memory (\(\altforgetparamabbrev = \egoforgetparamabbrev = 0.99\), corresponding to a characteristic time scale of approximately 100 steps). Our main observations are:
\begin{enumerate}

\item \(\knowledgeparam{X}=0.99; \expectationparam{X}=0; 
            \uncertaintyparam{X}=0; \randomparam{X}=0.01\). 
The responses of each
agent are essentially determined by its response disposition. The asymptotic behavior is given
by \(\genericprobjointnoargs{\infty}{A}{B}\) and \(\genericprobjointnoargs{\infty}{B}{A}\), the first modes  of the matrices \(\responsedispcondnoargs{A}\responsedispcondnoargs{B}\) and \(\responsedispcondnoargs{B}\responsedispcondnoargs{A}\), respectively. 
\figref{FigConv}
shows indeed the convergence to this state. The asymptotic values of the alter and ego memories are given
by \(\responsedispcondnoargs{A}\genericprobjointnoargs{\infty}{B}{A}\) and \(\responsedispcondnoargs{B}\genericprobjointnoargs{\infty}{A}{B}\)  The memory matrices have 
a ``blockwise'' structure with a block of maximal probability corresponding to the stimuli known by both agents.
 
\item \(\knowledgeparam{X}=0; \expectationparam{X}=0.99;
	 \uncertaintyparam{X}=0; \randomparam{X}=0.01\).
The response of each
agent is essentially determined by its ego memory. As expected,
the maximal variance is observed for this case. Indeed, there are long transient corresponding
to metastable states and there are many such states.

\item \(\knowledgeparam{X}=0;\expectationparam{X}=0;\uncertaintyparam{X}=0.99;\randomparam{X}=0.01\).
The response of each
agent is essentially determined by its alter memory, and the probability is higher for selecting messages that
induce responses with high uncertainty. Again, the asymptotic state is the state 
of maximal entropy.

\item \(\knowledgeparam{X}=0.99; \expectationparam{X}=0.99;
	 \uncertaintyparam{X}=-0.99; \randomparam{X}=0.01\). 
The response of each
agent is essentially determined by its alter memory, but the probability is lower for selecting messages that
induce responses with high uncertainty. The asymptotic memories
have an interesting structure. The alter and ego memories have block structures like that of both agents' response dispositions, but translated in the set of messages exchanged (contrast
\figref{FigDisp} and the bottom row of \figref{FMatrAs4}). The effect is clear if we keep in mind that, for \agent{A}'s ego memory,
the stimulus is always given by \agent{B}, while for \agent{A}'s alter memory, the stimulus is always given by \agent{A}. 

\end{enumerate}

Finally, we  note that, in this example, the asymptotic alter memory does not become
identical to the other agent's response disposition. Indeed, the perception that \agent{A} has from \agent{B} is biased by the fact that she chooses always the stimuli
according to her \textit{own} preferences (described in the appendix in more mathematical
terms). A prominent illustration of this is given in \figref{FMatrAs4} (bottom row).

These simulations shows how the final (asymptotic) outcome of the interaction may
differ when the parameters \(\genericcoeftrunc{i}\) are set to different, extremal values.
The role of the forgetting parameter is also important. In some sense, the asymptotic
matrices whose structure are the closest to the block structure of the response dispositions 
are the matrices in the bottom row of  \figref{FMatrAs4}, 
where the agent are able to ``forget.'' There is a higher contrast between the blocks
seen in the corresponding row of  \figref{FMatrAs2} when \(\altforgetparamabbrev = \egoforgetparamabbrev = 1\).  





\subsubsection{Behavior dependence on the parameters $\knowledgeparamtrunc, \expectationparamtrunc, \uncertaintyparamtrunc, \randomparamtrunc$.}
\label{sec:statbehavior}

Based on theoretical (\appendref{sec:transients}) and numerical 
(\secref{sec:complementaryknowledge}) considerations,
one expects changes 
in the asymptotic regime as the parameters \(\genericcoef{i}{A}\) and \(\genericcoef{i}{B}\) vary.
Note, however, that the general situation where \agentsxandy{A}{B} have different coefficients
requires investigations in an 8 dimensional space. Even if we impose constraints of the form in \eqnref{eq:paramconstraint},
this only reduces the parameter space to 6 dimensions.
A full investigation of the space of parameters is therefore beyond the scope of this paper and will
be done in a separate work. 
To produce a manageable parameter space, we focus here on the situation 
where  \agentsxandy{A}{B}
have identical coefficients and where \(\randomparamtrunc\) is fixed to a 
small value (\(0.01\)). 
The number of messages exchanged by the agents is \(\nummessages=11\). 

To explore the reduced, 
\(\knowledgeparamtrunc\expectationparamtrunc\) space,
we simulate interaction  with response dispositions of both agents
selected randomly (see \secref{sec:agenttypes} for details).
We conduct 10000 successive interaction sequences of length 11 for each 
pair of agents. In each sequence, the agent that transmit the first message is selected at random. Therefore we have a total exchange of 110000 messages.

We compute the asymptotic value of the joint entropy of the memories. We average the entropy over 100 realizations of the response dispositions. This allows us to 
present, \eg, the entropy of the alter memory as a function of 
\(\knowledgeparamtrunc\) and \(\expectationparamtrunc\).
 
In \figref{Entropy3D},
we show the joint entropy of the alter memory (the joint entropy of the ego memory is similar). The parameters \(\knowledgeparamtrunc\) and \(\expectationparamtrunc\) both vary in the interval \([0, 0.99]\).
Consequently, \(\uncertaintyparamtrunc \in [-0.99:0.99]\). There is an abrupt change in the entropy close to the line
\(\knowledgeparamtrunc+\expectationparamtrunc = 0.01\) (\(\uncertaintyparamtrunc=0\)). The transition does not occur exactly at said line, but 
depends on \(\knowledgeparamtrunc\) and \(\expectationparamtrunc\).

The surface in \figref{Entropy3D} shows a drastic change in the unpredictability of the pairs, and thus in the complexity of interactions,
when crossing a critical line in the \(\knowledgeparamtrunc\expectationparamtrunc\) parameter space. Note, however, that the transition
is not precisely at \(\uncertaintyparamtrunc=0\), 
but rather to a more complex line as discussed in the appendix.


\subsubsection{Rates of change.} \label{sec:learnrates}

While the asymptotically stable states of the memories are useful in understanding interaction, they are not the only important factors. Additionally, the
rates at which the memories change can be important. In particular, an inferior result quickly obtained may be more desirable than a superior result reached later. 

To investigate rates, we simulate two distinct types of agents exchanging \(\nummessages=5\) messages. The first 
type of agent is a ``teacher'' \agent{A} that has relatively simple behavior. Its response disposition has a banded structure, shown in \figref{fig:bandeddisp}. The teacher agent has a fixed behavior dependent only upon its response disposition (\(\expectationparam{A} = \uncertaintyparam{A}
= \randomparam{A} = 0\)). The values of \(\altforgetparam{A}\) and \(\egoforgetparam{A}\) are irrelevant, since the alter and ego memories contribute nothing.


The second type is a ``student'' \agent{B} that has more general behavior. The student agent has a random response disposition, as described in \secref{sec:agenttypes}. 
In general, the student agent uses all of its memories to determine the transition probability, and we vary the values of \(\knowledgeparam{B}, \expectationparam{B}, \uncertaintyparam{B},\) and \(\randomparam{B}\) to investigate the effects of the various parameters. We take the memories to be infinite, \(\altforgetparam{B} = \egoforgetparam{B} = 1\).

To assess the simulations, we focus on how rapidly the student agent is able to capture the behavior of the teacher agent in its memories. It is natural to determine this using the difference between the response disposition of the teacher agent and the alter memory of the student agent.  
We measure the difference using the 
distance~\(\seqdistance{\responsedispjointnoargs{A}}{\altmemjointnoargs{t}{B}{A}}\), defined by
\begin{equation}
    \sqseqdistance{\responsedispjointnoargs{A}}{\altmemjointnoargs{t}{B}{A}} = \half
        \sum_{r,s} \left(\responsedispjoint{A}{r}{s} - \altmemjoint{t}{B}{A}{r}{s}\right)^{2}
    \mathperiod
    \label{eq:testrefdistance}
\end{equation}
The value of \(\seqdistance{\responsedispjointnoargs{A}}{\altmemjointnoargs{t}{B}{A}}\) always lies in the 
interval~\([0, 1]\).

The distance defined by \eqnref{eq:testrefdistance}  requires memories from both agents, and thus cannot actually be calculated by either agent. However, by using the identity given by \eqnref{eq:memsymmetry}, we see 
that \(\sqseqdistance{\responsedispjointnoargs{A}}{\altmemjointnoargs{t}{B}{A}} 
= \sqseqdistance{\responsedispjointnoargs{A}}{\egomemjointnoargs{t}{A}{B}}\), which depends only on
memories of \agent{A} and is therefore in principle available to \agent{A}. This reveals an interesting asymmetry in the system; the teacher \agent{A} has an ability to assess the interaction that the student \agent{B} lacks. 

A simple way to determine a rate of change from the memories is to base it on the number of message exchanges it takes for the distance to first become smaller than some target distance. 
If \(N\) messages are exchanged before the target distance is reached, the rate is simply \(1/N\). The simulated interaction can be limited to a finite number of exchanges with the  rate set to zero if the target is not reached during the simulation. 

\begin{figure}[htbp]
	\centering
	\subfigure[Target distance \(0.10\)]{
		\includegraphics[width=\textwidth]{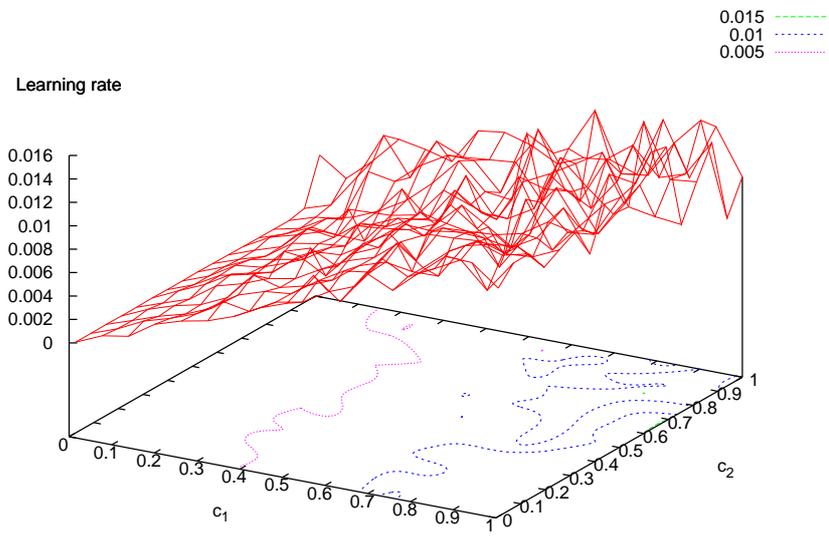}
		\label{fig:rate05target}}
	\subfigure[Target distance \(0.05\)]{
		\includegraphics[width=\textwidth]{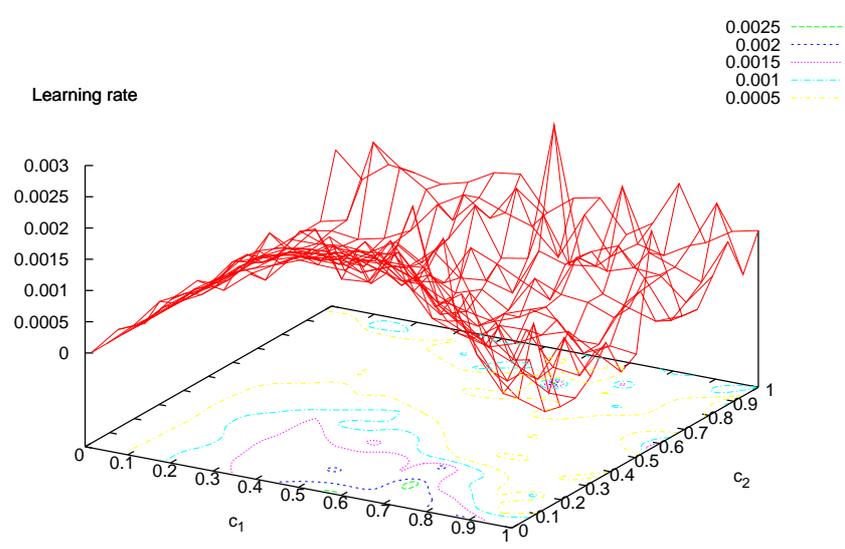}
		\label{fig:rate10target}}
	\caption{Rates of change for the alter memory of the student agent
	approaching the response disposition of the teacher agent. The colored lines in 
	the $\knowledgeparamtrunc$-$\expectationparamtrunc$ plane
	are level lines for the rates.}
	\label{fig:rates}
\end{figure}

In \figref{fig:rates}, we show the rates for different distances, averaged over 50 samples. In \figref{fig:rate05target}, the rate determined from a target distance of \(0.1\) is shown. For this case, the rates have a simple structure, generally proceeding more rapidly with \(\uncertaintyparam{B} < 0\). In contrast, \figref{fig:rate10target} presents quite different results based on a target distance of \(0.05\), with more complex structure to the rates. In this latter case, the highest rates occur with \(\uncertaintyparam{B} > 0\). We note that inclusion of the response disposition is crucial in this scenario. In particular, the rate of change goes to zero if the student agent suppresses its response disposition.


\section{Outlook.}\label{sec:outlook}

%
%
%

\subsection{Assessment of interactions.} \label{sec:assess}

In \secref{sec:memories}, we presented rules for updating the ego and alter
memories of an agent. These memories are updated during the course of
interactions in which the agent is involved. In contrast, the
response disposition is held fixed during interaction. In this section, we present a scheme for update of the response disposition, occurring intermittently at the end of interaction sequences.%
\footnote{For long interaction sequences, the
response disposition could be updated periodically---but less
frequently than the ego and alter memories---during the  
sequence.}

Thus far, the response disposition \(\responsedispjoint{A}{r}{s}\) has been presented as time independent. To generalize this, we number the interaction sequences and add an index to  the response disposition to indicate the sequence number, giving \(\timedeprespdisp{T}{A}\jointargs{r}{s}\).

Not all interactions are successful. For
a two-party interaction, it is possible 
that either or both of the parties will gain nothing, or that both benefit. We can incorporate this into the
model by introducing an assessment process  
for allowing updates to the response dispositions of the agents. 

To begin the assessment process, we can first calculate a joint probability distribution representative of the message sequence that was constructed during the interaction. Assuming the message sequence consists of \(k\)~messages labeled
 \(m_{1}, m_{2}, \ldots, m_{k-1}, m_{k}\), we find \(\seqempfreq{T}\), the empirical frequency of message pairs for interaction sequence~\(T\), using
\begin{equation}
    \seqempfreq{T}\jointargs{r}{s} = \frac{1}{k-1} \sum_{i=2}^{k} \krondelta{r}{m_{i}} \krondelta{s}{m_{i-1}}
    \label{eq:seqempfreqdef}
\end{equation}
for all \(r\) and \(s\). Note that \eqnref{eq:seqempfreqdef} utilizes all of the messages
exchanged, regardless of whether the agent ego and alter memories have infinite memories (\ie, 
\(\altforgetparam{A} = \egoforgetparam{A} = 1\))
or undergo ``forgetting'' (\ie, \(\altforgetparam{A} < 1\) 
or \(\egoforgetparam{A} < 1\)). This is due to an assumed period
of reflection in which details of the interaction can be considered at
greater length.

To actually make the comparison, we find the distance between the behavior 
expressed in the interaction sequence (using \(\seqempfreq{T}\)) and that expressed 
by the response disposition of the agents. A small distance is indicative of
relevance or usefulness of the interaction sequence to the agent, while a great 
distance suggests that the interaction sequence is, \eg, purely speculative or 
impractical.

Focusing on \agent{A} (a similar procedure is followed for \agent{B}), we calculate
the distance \(\seqdistance{\seqempfreq{T}}{\timedeprespdisp{T}{A}}\) using
\begin{equation}
    \sqseqdistance{\seqempfreq{T}}{\timedeprespdisp{T}{A}} = \half
        \sum_{r,s} \left(\seqempfreq{T}\jointargs{r}{s} - \timedeprespdisp{T}{A}\jointargs{r}{s}\right)^{2}
    \mathperiod
    \label{eq:distancedef}
\end{equation}
Since \(\seqempfreq{T}\) and \(\timedeprespdisp{T}{A}\) are both probability distributions, 
the distance \(\seqdistance{\seqempfreq{T}}{\timedeprespdisp{T}{A}}\) must lie in the 
interval \([0,1]\) The value of the distance must be below an acceptance 
threshold~\(\seqdistancethresh{A}\)
that reflects the absorbative capacity of \agent{A}, limiting what interaction 
sequences \agent{A} accepts.

The above considerations are used to define an update rule for the response 
disposition.
Formally, this is
\begin{equation}
    \timedeprespdisp{T}{A} = 
    		\left(1 - \sequpdaterate{A}\right) \timedeprespdisp{T-1}{A} 
				+ \sequpdaterate{A} \seqempfreq{T-1} 
    \mathcomma
    \label{eq:respdispupdaterule}
\end{equation}
where the update rate \(\sequpdaterate{A}\) is in the interval \([0,1]\).
The update rule in \eqnref{eq:respdispupdaterule} is applied if and only if 
\(\seqdistance{\seqempfreq{T}}{\timedeprespdisp{T}{A}} < \seqdistancethresh{A}\).  

\subsection{Communities of agents.} \label{sec:multiagent}

In this work, we have focused on interactions between pairs of agents. The agent definitions are posed in a manner that generalizes readily to larger groups of agents. However, some extensions to the agents described in \secref{sec:agents} are needed, in order to introduce rules by which agents select communication partners. 

A reasonable approach is to base rules for selecting interaction partners on affinities that the agents have for each other.  Letting \(\affinity{X}{Y}\) be the affinity that \agent{X} has for \agent{Y}, we set the probability that \agent{X} and \agent{Y} interact to be proportional to \(\affinityprod{X}{Y}\) (the product is appropriate because the affinities need not be symmetric). A pair of interacting agents  can then be chosen based on the combined affinities of all the agents present in the system.

As a starting point, the affinities can be assigned predetermined values, defining a fixed network of agents. More interestingly, the affinities can be determined based on the agent memories, so that interaction produces a dynamically changing network. A natural approach to take is to compare the behavior of an \agent{A} with its expectation of the behavior of another \agent{B}, as represented by the alter memory \(\altmemjointnoargs{t}{A}{B}\). For example, agents can be constructed to predominately interact with other agents they assess as similar by defining
\begin{equation}
	\affinity{X}{Y} = \frac{\affinityconst}{\affinityconst + \seqdistancesymb}
	\mathcomma
\end{equation}
where \(\affinityconst\) is a settable parameter and \(\seqdistancesymb\) is the distance defined in \eqnref{eq:testrefdistance}. The affinities can be viewed as the weights in a graph
describing the agent network. 
%

The outcomes of interactions  will depend not only on agent 
preferences and strategies, but also on the patterns of relations---encoded by the affinity 
graph---within which the agents are embedded. The structure of the network may play an 
important role in influencing the global spread of information or facilitating collective action. The interplay between network architecture and dynamical consequences is not 
straightforward, since the properties of the graphs that are relevant to individual and collective 
behavior of the agents will depend on the nature of the dynamical process describing the 
interactions. This process can be related to prior studies of contagion \citep[see, \eg,][]{BlaKruKruMar:2005}. However, unlike
classical epidemiological models, which assume spread of contagion to be a memory-free process, agents  are affected by both past and present interactions amongst one another. The probability of ``infection'' can exhibit threshold-like behavior, with the probability of adopting some behavioral pattern changing suddenly after some finite number of exposures.

\section{Summary} \label{sec:summary}

In this work, we have introduced an agent-based model of interaction. The model draws on concepts from Luhmann's contingency approach, and
builds on an earlier model by \citeauthor*{DitKroBan:2003}.
Agent interactions are 
formulated as the exchange of distinct messages that are selected based on individual
agent properties and the interaction history.

We expressed the model mathematically, in a form suitable for handling an arbitrary number 
of agents. Agent memories are formulated using a number of  
probability distributions. Two types of memories, the ego and alter memories, are
similar to the memories used by \citeauthor{DitKroBan:2003}, but we also introduced a 
response disposition showing individual  preferences for the agents. 
A selection probability for messages is calculated from the memories, along with a randomizing term  allowing for  invention. 

The selection probabilities are used to determine which messages the agents exchange, but the selection probabilities themselves depend on the ego and alter memories, which record the messages exchanged.  To analyze the properties of the model, we focused on pairs of agents. Using analytic methods and computational simulations, we explored the asymptotic properties of the agent memories and selection probability as the number of messages exchanged becomes very large, identifying some conditions for the existence and uniqueness of stable asymptotic solutions. Additionally, we investigated numerically the rates of change in shorter interaction sequences.

Finally, we sketched out intended extensions of the model. We described how the response dispositions of the agents might be updated and suggested how to handle larger groups of agents. In particular, these two extensions appear sufficient for investigating how inventions can be either adopted or rejected by the community of agents.
Since an agent uses its response disposition  with any of the other agents, 
the response disposition serves as a sort of general behavioral preference, in the form of messages
and transitions between messages. 
By tracking changes to the response disposition, we can distinguish between successful and unsuccessful inventions. In networks of agents, 
this approach would allow us to investigate how various network 
structures might facilitate or hinder the spread of inventions. This will be the subject of future research.

%

\section*{Acknowledgments.}

We are grateful to G.~Maillard for helpful references and comments on the mathematical aspects.
In addition we thank T.~Kr\"uger and W.~Czerny for many useful discussions.
We would like to acknowledge support from the Funda\c{c}\~ao para a Ci\^encia e a Tecnologia under Bolsa de Investiga\c{c}\~ao SFRH/BPD/9417/2002 (MJB) and Plurianual CCM (MJB, LS), as well as support from FEDER/POCTI, project 40706/MAT/2001 (LS).
Ph.~Blanchard, 
E.~Buchinger and B.~Cessac warmly acknowledge the CCM for its hospitality.

\appendix
\section{Mathematical analysis.} \label{sec:mathanalysisfull}

In this appendix, we develop the mathematical aspects
summarized in \secref{sec:mathsummary}. The formalism presented
here deals with the two-agent case, but some aspects generalize
to the multi-agent case. We are mainly interested in the asymptotic behavior of the model,
where the interaction between the two agents is infinitely long. Though
this limit is not reached in concrete situations, it can give us some idea of the system behavior for sufficiently long sequences. The asymptotic results essentially apply to the  finite  case, provided
that  the length
of the  sequence is longer than the largest characteristic time
for the transients (given by the spectral gap; see \appendref{sec:transients}).

\subsection{Basic statements.}

Recall that \(\selectprobcond{t}{A}{B}{r}{s}\) is
the conditional probability that \agent{A} selects message~\(r\) at time~\(t\) given that \agent{B} selected \(s\) at time~\(t-1\). It is given by   \eqnref{eq:transitionprobability},
\begin{equation} \label{UXY}
\selectprobcond{t}{A}{B}{r}{s}=\frac{1}{\normfactorcond{t}{A}{B}} \left( \knowledgeparam{A} \responsedispcond{A}{r}{s} + \expectationparam{A} \egomemcond{t}{A}{B}{r}{s} + 
\uncertaintyparam{A} \respentropy{t}{A}{B}{r} + \frac{\randomparam{A}}{\nummessages} \right)
\mathcomma
\end{equation}
where \(\respentropy{t}{A}{B}{r}\) is
\begin{equation}
\respentropy{t}{A}{B}{r} = -\sum_{r'=1}^{\nummessages} \altmemcond{t}{A}{B}{\rprime}{r}\log_{\nummessages} \left(\altmemcond{t}{A}{B}{\rprime}{r}\right)
\mathperiod
\end{equation}
Note that this term does not depend on \(s\).
The normalization factor \(\normfactorcond{t}{A}{B}\)
is given by
\begin{equation} \label{N} 
\normfactorcond{t}{A}{B}=1+ \uncertaintyparam{A} \left(\sum_{r=1}^{\nummessages}\respentropy{t}{A}{B}{r}-1 \right)
\end{equation}
and does not depend on time when \(\uncertaintyparam{A}=0\). Moreover, 
\begin{equation}\label{BoundsN}
	1-\uncertaintyparam{A} \leq 
	\normfactorcond{t}{A}{B} \leq 
	1+ \uncertaintyparam{A}  \left(\nummessages-1\right)
	\mathperiod
\end{equation}
It will be useful in the following to write \(\selectprobcond{t}{A}{B}{r}{s}\) in matrix 
form, so that
\begin{equation} \label{Umat}
\selectprobcondmatrix{t}{A}{B}=\frac{1}{\normfactorcond{t}{A}{B}}
	\left(\knowledgeparam{A} \responsedispcondmatrix{A} +
	\expectationparam{A} \egomemcondmatrix{t}{A}{B} + 
	\uncertaintyparam{A} \respentropymatrix{t}{A}{B} + 
	\randomparam{A} \normalizeduniformprobmatrix \right)
	\mathcomma
	\label{eq:matrixselectprob}
\end{equation}
with a corresponding equation for the \(\selectprobcondmatrix{t}{B}{A}\).
The matrix \(\normalizeduniformprobmatrix\) is the uniform conditional probability, with the form
\begin{equation}
\normalizeduniformprobmatrix =
\frac{1}{\nummessages} 
\left(
\begin{array}{cccccc}
1& \cdots &1\\
\vdots&\ddots&\vdots\\
1& \cdots &1
\end{array}
 \right)
 \mathperiod
\end{equation} 
We use the notation \(\respentropymatrix{t}{A}{B}\) for the matrix
with the \(r,s\) element given by  \(\respentropy{t}{A}{B}{r}\).

For simplicity, we assume here that \agent{B} always starts 
the interaction.
This means that \agent{B} selects the message at odd times \(t\) and \agent{A} selects 
at even times. 
When it does not harm the clarity of the expressions, we will drop the subscripts labeling the agents from the transition probabilities and call \(\selectprobcondmatrixtrunc{t}\) the transition matrix at time \(t\), with the convention that
\(\selectprobcondmatrixtrunc{2n} \equiv  \selectprobcondmatrix{2n}{A}{B}\) and   
\(\selectprobcondmatrixtrunc{2n+1} \equiv  \selectprobcondmatrix{2n+1}{B}{A}\). 
Where possible, we will further simplify the presentation by focusing on just one of the agents, with the argument for the other determined by a straightforward exchange of the roles of the agents.

\subsection{Dynamics.}

Call \(\genericprobmargabbrev{t}\margargs{r}\) the probability that the relevant agent selects 
message~\(r\) at time \(t\).
The evolution of \(\genericprobmargabbrev{t}\margargs{r}\) is determined by a 
one parameter family of 
(time dependent) transition  matrices, like in Markov chains. However, the evolution is not Markovian
since the transition matrices depend on the entire past history (see \secref{sec:existunique}).

Denote by \(\namedinfinitemessagesequence = \infinitemessagesequence{1}{2}{t-1}{t}\)
an infinite
message exchange sequence where \(\messagechoice{t}\) is the \(t\)-th exchanged symbol.
The probability that the finite subsequence 
\(\messagesequence{1}{2}{t-1}{t}\)
has been selected 
is given by
\begin{eqnarray}
	\lefteqn{\mbox{Prob}\left[\messagechoice{t}, \messagechoice{t-1}, \dots, 
				\messagechoice{2},\messagechoice{1} \right] = } \nonumber \\
	& &	\selectprobcondtrunc{t}\condargs{\messagechoice{t}}{\messagechoice{t-1}}
		\selectprobcondtrunc{t-1}\condargs{\messagechoice{t-1}}{\messagechoice{t-2}} 
		\dots 
		\selectprobcondtrunc{2}\condargs{\messagechoice{2}}{\messagechoice{1}}
		\genericprobmargabbrev{1}\margargs{\messagechoice{1}}
\mathperiod
\end{eqnarray}
Thus,
\begin{equation} 
	\genericprobmargmatrixtrunc{t} = 	\selectprobcondmatrixtrunc{t} 
							\selectprobcondmatrixtrunc{t-1}
							\selectprobcondmatrixtrunc{t-2} 
							\cdots 
							\selectprobcondmatrixtrunc{2}
							\genericprobmargmatrixtrunc{1}
	\mathperiod
	\label{Pt}
\end{equation}
Since \agent{B} selects the initial message, we have  
\(\genericprobmargabbrev{1}\margargs{r} = 
\sum_{s=1}^{\nummessages} \responsedispjoint{B}{r}{s}\).
As well, the joint probability of the sequential messages \(\messagechoice{t}\) and \(\messagechoice{t-1}\) is
\begin{equation} \label{Pairt}
\mbox{Prob}\left[\messagechoice{t}, \messagechoice{t-1}\right] = \selectprobcondtrunc{t}(\messagechoice{t}\given\messagechoice{t-1}) \genericprobmargabbrev{t-1}\left(\messagechoice{t-1} \right)
\mathperiod
\end{equation}

Note that the \(\selectprobcondmatrixtrunc{t}\) depend on the response dispositions.
 Therefore one can at most expect statistical statements referring to some probability measure 
on the space of response disposition matrices (recall that the selection probability for the 
first message
is also determined by the response disposition). We therefore need to consider statistical
averages to characterize the typical behavior of the model for fixed values of the coefficients~\(\genericcoef{i}{A}\) and~\(\genericcoef{i}{B}\).

\subsection{Existence and uniqueness of a stationary state.} \label{sec:existunique}

In this section, we briefly discuss the asymptotic properties of the ME model considered as a 
stochastic process. 
The main objective is to show the existence and uniqueness
of a stationary state, given that the probability for a given message to be selected at time \(t\) depends on the entire past
(via the ego and alter memories). In the case where \(c_2=c_3=0\), the ME model is basically
a Markov process and it can be handled with the standard results in the field.

In the general case, one has first to construct the probability  on the set of infinite
trajectories \({\cal A}^{{\sf I\!N}}\) where \({\cal A} = \set{1, 2, \ldots n_S}\) is the 
set of messages. 
The corresponding sigma algebra
\({\cal F}\) is constructed as usual with the cylinder sets. In the ME model the transition probabilities corresponding
to a given agent are determined via the knowledge of the sequence of message pairs (second order approach), but
the following construction holds for an \(n\)-th order approach, \(n \geq 1\), where the transition probabilities
depends on a sequence of \(n\)-tuples. 
If \(\messagesequence{1}{2}{t-1}{t}\) is a message sequence,
we denote by
\(\omega_1, \omega_2, \ldots \omega_k \) the sequence of
pairs \( \omega_{1}= \messagechoicepair{1}{2}, \omega_{2} = \messagechoicepair{3}{4},
\ldots \omega_{k} = \messagechoicepair{2k-1}{2k}\).
Denote by \(\omega_l^m = \left(\omega_l \dots \omega_m \right)\), where \(m>l\).
We construct a family of conditional probabilities given by
\begin{equation}
	P\left[\omega_t=(r,s) |\omega_1^{t-1} \right] = 
		c_1 Q(r,s) + 
		c_2  {\cal E}\left[\omega_t=(r,s) |\omega_1^{t-1} \right] + 
		c_3 H\left(A\left[\cdot \given \omega_1^{t-1} \right]\right)(r)+
		c_4
\end{equation}
for \(t>1\), where 
\begin{equation}
{\cal E}\left[\omega_t=(r,s) | \omega^{t-1}_1\right]= \frac{1}{t} \sum_{k=1}^{t-1} \chi(\omega_k =(r,s))
\mathcomma
\end{equation}
with \(\chi()\) being the indicatrix function (\(A\left[\cdot \given \omega_1^{t-1}\right]\) has the same form, see \secref{sec:memories}).
 For \(t=1\) the initial pair \(\omega_1\) is drawn using the response disposition
as described in the text. Call the corresponding initial probability \(\mu\). 

We have 
\begin{equation}
	P\left[\omega^m_{n+1}|\omega^n_1 \right] =
	P\left[\omega_{n+1}|\omega^n_1 \right]
	P\left[\omega_{n+2}|\omega^{n+1}_1 \right]
	\dots
	P\left[\omega_m|\omega^{m-1}_1 \right]
\end{equation}
and, \(\forall m,l >1\),
\begin{equation}
\sum_{\omega^{m-1}_1} P\left[\omega^{m-1}_1|\omega_1 \right] P\left[\omega^{m+l}_m|\omega^{m-1}_1 \right]
=P\left[\omega^{m+l}_m|\omega_1 \right]
\mathperiod
\end{equation}
From this relation, we can define a probability on the cylinders \(\left\{ \omega^{m+l}_m\right\}\) by
\begin{equation}
	\label{ProbaCyl}
	P\left[\omega^{m+l}_m\right]=\sum_{\omega_1}
	P\left[\omega^{m+l}_m|\omega_1 \right]\mu(\omega_1)
	\mathperiod
\end{equation}
This measure extends then on the space of trajectories \(\left({\cal A}^{{\sf I\!N}},
{\cal F}\right)\)
by Kolmogorov's theorem. It depends on the probability \(\mu\) of choice for the first symbol (hence
it depends on the response disposition of the starting agent). A stationary state is then a  shift invariant measure 
on the set of infinite sequences. We have not yet been able to find rigorous conditions ensuring the existence
of a  stationary state, but some arguments are given below. In the following, we assume this existence. 

Uniqueness is expected from standard ergodic argument whenever all transitions 
are permitted. This is the case when both
\(\randomparamtrunc > 0\) and \(\uncertaintyparamtrunc \geq 0\)  
(the case where \(\uncertaintyparamtrunc\) is negative is trickier and is discussed below).
When the invariant state is unique, it is ergodic and in this case the empirical frequencies corresponding to alter and ego
memories converge (almost surely) to a limit corresponding to the marginal  probability \(P[\omega]\) obtained from \eqnref{ProbaCyl}.
Thus, the transition matrices \(\selectprobcondmatrixtrunc{t}\) also converge to a limit.

In our model, uniqueness basically implies that  the asymptotic 
behavior of the agent does not
depend on the past. This convergence can be interpreted
as a stabilization of the behavior of each agent when interacting with the other. 
After a sufficiently long
exchange of messages, each agent has a representation of the other that does not evolve in time (alter memory).
It may evaluate the probability of all possible reactions to the possible stimuli, and this probability is not modified by further exchanges. 
Its own behavior, and the corresponding representation (ego memory), is similarly fixed. 

In the next section we derive  explicit solutions for the asymptotic alter and ego memories. We are able to identify
several regions in the parameters space where the solution is unique. But it is not excluded that several solutions
exist in other regions, especially for \(\uncertaintyparamtrunc < 0\). In this case, the long time behavior of the agent depends
on the initial conditions. A thorough investigation of this point is, however, beyond the scope of this paper.

Note that the kind of processes encountered in the ME model  has interesting relations
 with the so-called ``chains with complete connections''
\citep[see][]{Mai:2003} 
%
%
for a 
comprehensive introduction and detailed bibliography). There is in particular an interesting
relation between chains with complete connections and the Dobrushin-Landford-Ruelle construction of Gibbs measures \citep{Mai:2003}.
Note, however, that the transition probabilities in our model  do not obey the continuity property  usual in the context of 
chains with complete
connections. 

\subsection{Transients and asymptotics.} \label{sec:transients}

The time evolution and the asymptotic properties 
of \eqnref{Pt} are determined by the spectral properties of the
matrix product \(\selectprobcondmatrixtrunc{t}\selectprobcondmatrixtrunc{t-1}\selectprobcondmatrixtrunc{t-2} \dots \selectprobcondmatrixtrunc{2}\).
Assume that the \(\selectprobcondmatrixtrunc{t}\)  converge to a limit. Then, since the
\(\selectprobcondmatrixtrunc{t}\) are conditional probabilities, there exists, 
by the Perron-Frobenius theorem,
an eigenvector associated with an eigenvalue 1, corresponding to a stationary state (the states may be different for odd and even times). 
The stationary state is not necessarily unique.  It is unique if there exist a time \(t_0\)
such that for all \(t > t_0\)
the matrices \(\selectprobcondmatrixtrunc{t}\) are recurrent and aperiodic (ergodic). For \(\uncertaintyparamtrunc > 0\) 
and \(\randomparamtrunc > 0\), ergodicity is assured by the presence
of the matrix \(\normalizeduniformprobmatrix\), which allows transitions from any message to any other message.
This means that for positive \(\uncertaintyparamtrunc\) the asymptotic behavior of the agents does not depend
on the history (but the transients depend on it, and they can be very long). 

For \(\uncertaintyparamtrunc < 0\), it is possible that some transitions
allowed by the matrix \(\normalizeduniformprobmatrix\) are canceled and that the transition matrix \(\selectprobcondmatrixtrunc{t}\)
loses the recurrence property for sufficiently large \(t\). 
In such a case, it is not even guaranteed that a stationary regime exists.

The convergence rate of the process is determined by the spectrum of the 
\(\selectprobcondmatrixtrunc{t}\) and especially
by the spectral gap (distance between the largest eigenvalue of one 
and the second largest eigenvalue). Consequently, studying the statistical properties of 
the evolution for the spectrum of the \(\selectprobcondmatrixtrunc{t}\)
provides information such as the rate at which \agent{A}  has stabilized its behavior
when interacting with \agent{B}.  

The uncertainty term and the noise term play particular roles
in the evolution determined by \eqnref{Umat}.
First, the matrix \(\normalizeduniformprobmatrix\) entering in the noise term has 
eigenvalue 0 with multiplicity \(\nummessages-1\) 
and eigenvalue 1 with multiplicity 1. The eigenvector corresponding to the latter eigenvalue is 
\begin{equation}
	\uniformmatrixeigenvector = \frac{1}{\nummessages}
		\left( \begin{array}{c}
			1\\
			1\\
			\vdots\\
			1
		\end{array} \right)
	\mathcomma
\end{equation} 
which is the uniform probability vector, corresponding to maximal entropy.
Consequently, \(\normalizeduniformprobmatrix\) is a  projector onto \(\uniformmatrixeigenvector\). 

Second, the uncertainty term does not depend on the stimulus \(s\). It corresponds therefore to a matrix where all
the entries in a row  are equal. More precisely, set \(\alpha^{(t)}_r=\respentropy{t}{A}{B}{r}\).
Then one can write the corresponding matrix in the
form \(\nummessages \respentropymatrixeigenvectortrunc{t} 
\uniformmatrixeigenvector^{\mathrm{T}}\), where \(\respentropymatrixeigenvectortrunc{t}\) is the vector
\begin{equation}
	\respentropymatrixeigenvectortrunc{t} = 
		\left( \begin{array}{c}
			\alpha^{(t)}_1 \\
			\alpha^{(t)}_2 \\
			\vdots \\
			\alpha^{(t)}_{\nummessages} 
		\end{array} \right)
\end{equation} 
and \(\uniformmatrixeigenvector^{\mathrm{T}}\) is the transpose of 
\(\uniformmatrixeigenvector\).
It follows that the 
uncertainty term has a 0 eigenvalue with multiplicity \(n-1\)  and an eigenvalue \(\sum_{r=1}^{\nummessages}\respentropy{t}{A}{B}{r}\)
with corresponding eigenvector \(\respentropymatrixeigenvectortrunc{t}\). 
It is also apparent that, for any probability vector \(\genericprobmatrixsymb\), we have
\(\respentropymatrix{t}{A}{B} \genericprobmatrixsymb = 
\nummessages \respentropymatrixeigenvector{t}{A}
\uniformmatrixeigenvector^{\mathrm{T}} \genericprobmatrixsymb 
= \respentropymatrixeigenvector{t}{A}\).

The action of \(\selectprobcondmatrixtrunc{t}\) on \(\genericprobmargmatrixtrunc{t}\) is then given by
\begin{eqnarray}
\genericprobmargmatrix{t+1}{A} & = & 
	\selectprobcondmatrix{t}{A}{B}\genericprobmargmatrix{t}{B} \nonumber \\
	& = & \frac{1}{\normfactorcondabbrev{t}{A}{B}} \left (
		\left[ \knowledgeparam{A} \responsedispcondmatrix{A} 
			+ \expectationparam{A} \egomemcondmatrix{t}{A}{B} \right]
			\genericprobmargmatrix{t}{B}
		+ \uncertaintyparam{A} \respentropymatrixeigenvector{t}{A}
		+ \randomparam{A} \uniformmatrixeigenvector
		\right )
	\mathperiod
	\label{eq:transitionprobaction}
\end{eqnarray}
The expression in \eqnref{eq:transitionprobaction} warrants several remarks. 
Recall that all the vectors above have positive entries.
Therefore the noise term \(\randomparam{A} \uniformmatrixeigenvector\) tends
to ``push'' \(\genericprobmargmatrix{t}{B}\) in the direction of the
vector of maximal entropy, with the effect of increasing the entropy whatever the initial probability
and the value of the coefficients.
The uncertainty term \(\uncertaintyparam{A} \respentropymatrixeigenvector{t}{A}\) plays a 
somewhat similar role in the sense that it also has its image on a particular
vector. However, this vector is not static, instead depending on the evolution via the alter 
memory.  Further, the coefficient \(\uncertaintyparam{A}\) may have either a positive or 
a negative value. A positive \(\uncertaintyparam{A}\) increases the contribution of \(\respentropymatrixeigenvector{t}{A}\) but a negative \(\uncertaintyparam{A}\) decreases
the contribution. Consequently, we expect drastic changes in the model evolution when we change the sign of \(\uncertaintyparam{A}\)---see \secref{sec:simresults} and especially 
\figref{Entropy3D} for a striking demonstration of these changes.

\subsection{Equations of the stationary state.} \label{sec:stationarystateequations}

The influence of the coefficients \(\genericcoef{i}{A}\) and \(\genericcoef{i}{B}\) is 
easier to handle when 
an asymptotic state, not necessarily unique and possibly sample dependent, is reached. 
In this case, we must still distinguish between odd times (\agent{B} active) and even times (\agent{A} active), 
that is, \(\genericprobmargmatrixtrunc{t}\) has two accumulation points
depending on whether \(t\) is odd or even. 
Call  \(\genericprobmarg{\infty}{A}{\messagechoicesymbol}\)  the probability that,
in the stationary regime, \agent{A}  selects the message \(\messagechoicesymbol\) 
during interaction with \agent{B}. 
In the same way,  call \(\genericprobjoint{\infty}{A}{B}{r}{s}\)  
the asymptotic joint probability that \agent{A} responds with message \(r\)
to a stimulus message \(s\) from \agent{B}.  Note that, in general, 
\(\genericprobjoint{\infty}{A}{B}{r}{s} \neq \genericprobjoint{\infty}{B}{A}{r}{s}\), but
\(\genericprobmarg{\infty}{A}{\messagechoicesymbol} = \sum_{s}\genericprobjoint{\infty}{A}{B}{\messagechoicesymbol}{s} =
\sum_{r}\genericprobjoint{\infty}{B}{A}{r}{\messagechoicesymbol}\) and
\(\genericprobmarg{\infty}{B}{\messagechoicesymbol} = \sum_{s}\genericprobjoint{\infty}{B}{A}{\messagechoicesymbol}{s} =
\sum_{r}\genericprobjoint{\infty}{A}{B}{r}{\messagechoicesymbol}\) since each message 
simultaneous is both a stimulus and a response. 

Since alter and ego memories are empirical frequencies of stimulus-response pairs,
the convergence to a stationary state implies that \(\egomemjoint{2n}{A}{B}{r}{s}\) 
converges to a limit  \(\egomemjoint{\infty}{A}{B}{r}{s}\) 
which is precisely the asymptotic probability of stimulus-response pairs. 
Therefore, we have
\begin{eqnarray}
	\egomemjoint{\infty}{A}{B}{r}{s} &=& 
				\genericprobjoint{\infty}{A}{B}{r}{s}\label{EABPABa}\\
	\egomemjoint{\infty}{B}{A}{r}{s} &=&
				\genericprobjoint{\infty}{B}{A}{r}{s}\label{EABPABb}
	\mathperiod
\end{eqnarray}
Thus, the \(\selectprobcondmatrix{2n}{A}{B}\) converge to a limit \(\selectprobcondmatrix{\infty}{A}{B}\), where
\begin{equation}
	\selectprobcondmatrix{\infty}{A}{B} = \frac{1}{\normfactorcond{\infty}{A}{B}}
			\left( 	\knowledgeparam{A} \responsedispcondmatrix{A} + 
				\expectationparam{A} \egomemcondmatrix{\infty}{A}{B} + 
				\uncertaintyparam{A} \respentropymatrix{\infty}{A}{B} + 
				\randomparam{A}\normalizeduniformprobmatrix 
			\right)
	\mathperiod
	\label{CalcWXY}
\end{equation}
From \eqnref{Pairt}, we have  \(\genericprobmarg{\infty}{A}{r}=\sum_{s} \selectprobcond{\infty}{A}{B}{r}{s}\genericprobmarg{\infty}{B}{s}\).
Hence,
\begin{eqnarray}
	\genericprobmargmatrix{\infty}{A} &=& 
		\selectprobcondmatrix{\infty}{A}{B} \genericprobmargmatrix{\infty}{B}\\
	\genericprobmargmatrix{\infty}{B} &=& 
		\selectprobcondmatrix{\infty}{B}{A} \genericprobmargmatrix{\infty}{A}
\end{eqnarray}
and
\begin{eqnarray}
	\genericprobmargmatrix{\infty}{A} &=& 
		\selectprobcondmatrix{\infty}{A}{B} \selectprobcondmatrix{\infty}{B}{A} 
								\genericprobmargmatrix{\infty}{A}\\
	\genericprobmargmatrix{\infty}{B} &=&
		\selectprobcondmatrix{\infty}{B}{A} \selectprobcondmatrix{\infty}{A}{B}
								\genericprobmargmatrix{\infty}{B}	
	\mathperiod
\end{eqnarray}
It follows that the asymptotic probability \(\genericprobmargmatrix{\infty}{A}\) 
is an eigenvector of \(\selectprobcondmatrix{\infty}{A}{B} \selectprobcondmatrix{\infty}{B}{A}\) 
corresponding to the eigenvalue 1. We will call this eigenvector
the first mode of the corresponding matrix. Therefore,
the marginal ego memory of \agent{A} 
converges to the first mode of \(\selectprobcondmatrix{\infty}{A}{B}\selectprobcondmatrix{\infty}{B}{A}\). 
A numerical example is provided  in \secref{sec:complementaryknowledge}.

Combining \eqnxandy{Pairt}{EABPABa}, \(\genericprobjoint{\infty}{A}{B}{r}{s}=\selectprobcond{\infty}{A}{B}{r}{s}\genericprobmarg{\infty}{B}{s}=\egomemjoint{\infty}{A}{B}{r}{s}\). 
But \(\genericprobmarg{\infty}{B}{s}= \sum_{r}\genericprobjoint{\infty}{A}{B}{r}{s}= \sum_{r}\egomemjoint{\infty}{A}{B}{r}{s}\), so
\begin{eqnarray}
	\egomemcond{\infty}{A}{B}{r}{s}&=&\selectprobcond{\infty}{A}{B}{r}{s}
	\label{eq:asymptoticegoisselectprob}\\
	\egomemcond{\infty}{B}{A}{r}{s}&=&\selectprobcond{\infty}{B}{A}{r}{s}
	\mathperiod
\end{eqnarray}
Therefore, using the relation 
\(\egomemcond{\infty}{A}{B}{r}{s}=\altmemcond{\infty}{B}{A}{r}{s}\), we have
\begin{eqnarray}
	\egomemcond{\infty}{A}{B}{r}{s} &=& 
		\frac{1}{\normfactorcond{\infty}{A}{B}}
			\left[	\knowledgeparam{A} \responsedispcond{A}{r}{s} + 
				\expectationparam{A} \egomemcond{\infty}{A}{B}{r}{s} + 
				\uncertaintyparam{A} 
						\egorespentropy{\infty}{B}{A}{r} 
				+ \randomparam{A} 
			\right] \nonumber \\*
	\label{EABSola}\\
	\egomemcond{\infty}{B}{A}{r}{s} &=&
		\frac{1}{\normfactorcond{\infty}{B}{A}}
			\left[	\knowledgeparam{B} \responsedispcond{B}{r}{s} + 
				\expectationparam{B} \egomemcond{\infty}{B}{A}{r}{s} + 
				\uncertaintyparam{B}
						\egorespentropy{\infty}{A}{B}{r} 
				+ \randomparam{B} 
			\right]
	\mathperiod \nonumber \\
	\label{EABSolb}
\end{eqnarray}

\subsection{Solutions of the stationary equations.}

Define 
\begin{equation}
	\normalizedcoef{i}{A}= 
	\frac{\genericcoef{i}{A}}{\normfactorcond{\infty}{i}{A}-\expectationparam{A}} 
	\mathperiod
\end{equation}
With this definition, \eqnxandy{EABSola}{EABSolb} become:
\begin{eqnarray}
	\egomemcond{\infty}{A}{B}{r}{s}&=&
	\normalizedcoef{1}{A} \responsedispcond{A}{r}{s} + 
		\normalizedcoef{3}{A} 
			\egorespentropy{\infty}{B}{A}{r} + 
		\normalizedcoef{4}{A} 
	\label{eq:stationaryegoabcoupled} \\
	\egomemcond{\infty}{B}{A}{r}{s}&=&
	\normalizedcoef{1}{B} \responsedispcond{B}{r}{s} + 
		\normalizedcoef{3}{B} 
			\egorespentropy{\infty}{A}{B}{r} + 
		\normalizedcoef{4}{B}
	\label{eq:stationaryegobacoupled} 
	\mathperiod
\end{eqnarray}
We next plug \eqnref{eq:stationaryegobacoupled} into \eqnref{eq:stationaryegoabcoupled}. 
After some manipulation, we obtain
\begin{eqnarray}\label{EABSolFinal}
\lefteqn{\egomemcond{\infty}{A}{B}{r}{s} =} \nonumber \\
&&\normalizedcoef{1}{A} \responsedispcond{A}{r}{s} + 
	\normalizedcoef{4}{A} + 
	\normalizedcoef{3}{A} 
		\memoryentropy{\normalizedcoef{1}{B} \responsedispcond{B}{\cdot}{r} +
					\normalizedcoef{4}{B} } \nonumber \\
&& {} - \normalizedcoef{3}{A} \sum_{r'}(\normalizedcoef{1}{B} \responsedispcond{B}{\rprime}{r}+\normalizedcoef{4}{B})
\log_{\nummessages}\left(1+ \frac{\normalizedcoef{3}{B} \egorespentropy{\infty}{A}{B}{r'}}{\normalizedcoef{1}{B} \responsedispcond{B}{\rprime}{r}+\normalizedcoef{4}{B}}\right) \nonumber \\
&&{} -  \normalizedcoef{3}{A} \normalizedcoef{3}{B}\sum_{r'}\log_{\nummessages}\left(\normalizedcoef{1}{B} \responsedispcond{B}{\rprime}{r}+\normalizedcoef{4}{B} \right)\egorespentropy{\infty}{A}{B}{r'} \nonumber \\
&&{} -  \normalizedcoef{3}{A} \normalizedcoef{3}{B}\sum_{r'}\egorespentropy{\infty}{A}{B}{r'}
\log_{\nummessages}\left(1+ \frac{\normalizedcoef{3}{B} \egorespentropy{\infty}{A}{B}{r'}}{\normalizedcoef{1}{B} \responsedispcond{B}{\rprime}{r}+\normalizedcoef{4}{B}}\right)   \nonumber \\
\label{eq:hideousexpandedform}
\mathcomma
\end{eqnarray}
which uncouples the expression for 
\(\egomemcond{\infty}{A}{B}{r}{s}\) from that for
\(\egomemcond{\infty}{B}{A}{r}{s}\).
In some sense, \eqnref{eq:hideousexpandedform} provides a solution of the model
with two agents, since it captures the asymptotic behavior of the 
ego and alter memories (provided the stationary regime exists). 
However, the solution to \eqnref{eq:hideousexpandedform} is difficult
to obtain for the general case and 
depends on all the parameters \(\genericcoef{i}{A}\) and \(\genericcoef{i}{B}\). 
Below, we discuss a few specific situations.

This form has the advantage that it accommodates series expansion 
in \(\uncertaintyparam{A}\).
However, \(\normalizedcoef{3}{A}\) depends on \(\egomemcondnoargs{\infty}{A}{B}\)  via the normalization factor from \eqnref{N} and high order terms in the expansion are tricky to obtain.
Despite this, the bounds given in \eqnref{BoundsN} ensure that \(\normalizedcoef{3}{A}\) 
is small whenever \(\uncertaintyparam{A}\) is small. This allows us to characterize the 
behavior of the
model when \(\uncertaintyparam{A}\) changes its sign. This is of principle importance, since 
we shall see that a transition occurs near \(\uncertaintyparamtrunc=0\).
It is clear that the first term  in  \eqnref{EABSolFinal} is of order zero 
in \(\normalizedcoef{3}{A}\), the second term is  of order
one, and the remaining terms are  of higher order. 

When \(\uncertaintyparam{A}=0\), \agent{A} does not use its alter memory
in response selection. Its asymptotic ego memory is a simple function of its 
response disposition, with the form
\begin{equation}
	\egomemcond{\infty}{A}{B}{r}{s}=
	\normalizedcoef{1}{A} \responsedispcond{A}{r}{s} + \normalizedcoef{4}{A}
	\mathperiod
\end{equation}
When \(\uncertaintyparam{A}\) is small,  \egomemcond{\infty}{A}{B}{r}{s} becomes a 
nonlinear function of the conditional response disposition for \agent{B}, so that
\begin{equation}
	\egomemcond{\infty}{A}{B}{r}{s}=
	\normalizedcoef{1}{A} \responsedispcond{A}{r}{s} + 
	\normalizedcoef{4}{A} + 	
	\normalizedcoef{3}{A} \memoryentropy{\normalizedcoef{1}{B} 
		\responsedispcond{B}{\cdot}{r}+\normalizedcoef{4}{B} }
\end{equation}
An explicit, unique solution exists in this case. 

\begin{figure}[thbp]
	\centering
	\subfigure[Infinite memory ($\altforgetparamabbrev=\egoforgetparamabbrev=1$).]
		{\includegraphics[width=8cm,clip=false]{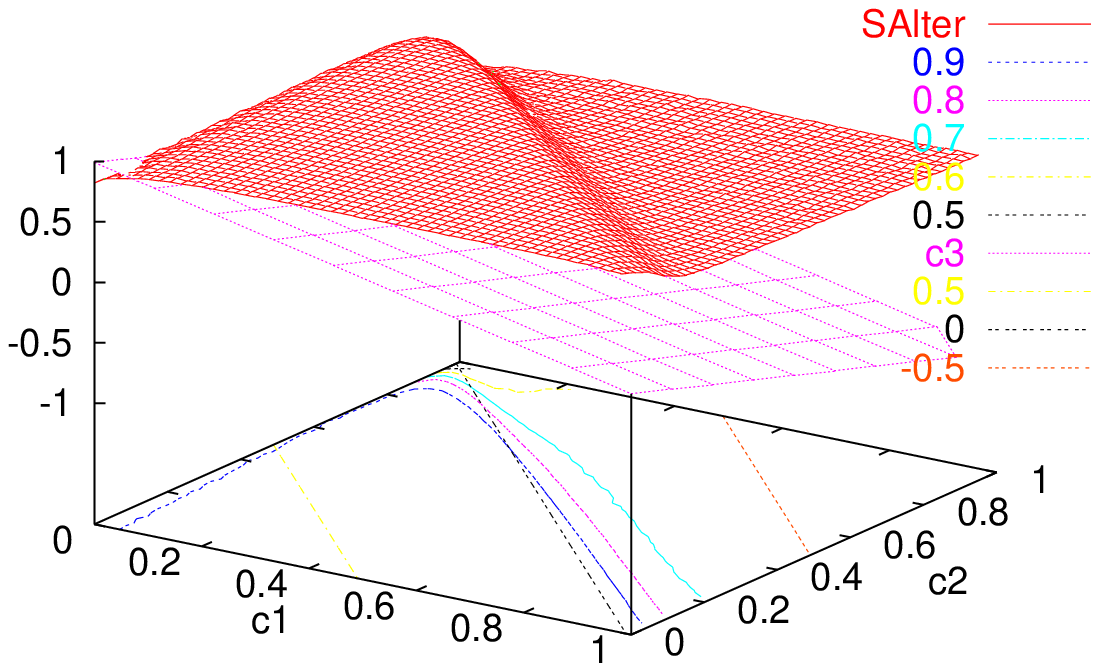}}
	\subfigure[Finite memory ($\altforgetparamabbrev=\egoforgetparamabbrev=0.99$).]
		{\includegraphics[width=8cm,clip=false]{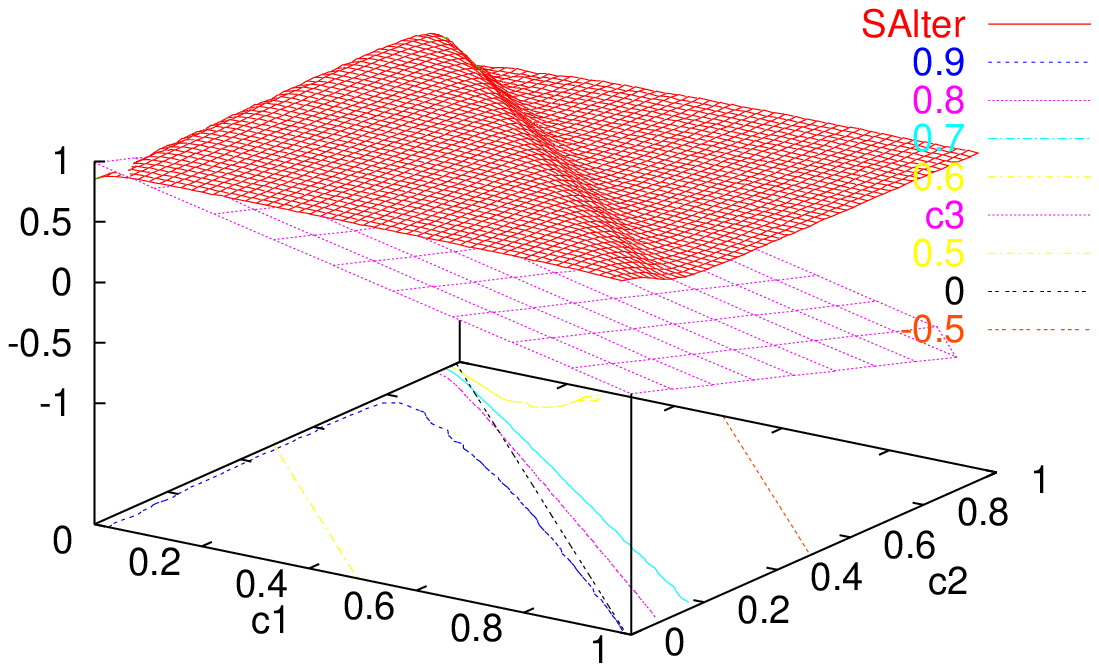}}
	\caption{Asymptotic joint entropy for  alter memory of \agent{A}, with (a) infinite 
	memories and (b) finite memories.
	The plane represents $\uncertaintyparamtrunc=0$. 
	The colored lines in the $\knowledgeparamtrunc$-$\expectationparamtrunc$ plane
	are level lines for the joint entropy, while the  black line shows where the
	$\uncertaintyparamtrunc=0$ plane intersects 
	the $\knowledgeparamtrunc$-$\expectationparamtrunc$ plane.
	}
	\label{Entropy3D}  
\end{figure}

For small values of \(\uncertaintyparam{A}\), the derivative of
\(\egomemcond{\infty}{A}{B}{r}{s}\) with respect to \(\uncertaintyparam{A}\) 
is proportional to 
\(\memoryentropy{\normalizedcoef{1}{B} \responsedispcond{B}{\cdot}{r} + 
\normalizedcoef{4}{B} }\). Note that the derivative depends nonlinearly on the 
coefficients \(\knowledgeparam{B}\) and \(\randomparam{B}\), and that, therefore, the level
lines \(\memoryentropy{\normalizedcoef{1}{B} \responsedispcond{B}{\cdot}{r} + 
\normalizedcoef{4}{B} }=C\) depend on \(\knowledgeparam{B}\) and \(\randomparam{B}\)
(see \figref{Entropy3D} where the levels lines are 
drawn---they do not coincide with $c_3=0$).
This slope can be steep if
the uncertainty in \agent{B}'s response disposition is high. For example, if  \(\knowledgeparam{B}\) is small, the entropy is high.
There exists therefore a transition, possibly sharp,
near \(\uncertaintyparam{A} = 0\). 

As \(\normalizedcoef{3}{A}\) further increases, we must deal with a more complex,
nonlinear equation for the ego memory. In the general case, several solutions may
exist.

A simple case corresponds to having \(\uncertaintyparam{A}=\uncertaintyparam{B}=1\). Indeed, in this case, we have
\begin{equation}
 	\egomemcond{\infty}{A}{B}{r}{s}=-  \normalizedcoef{3}{A}
	\normalizedcoef{3}{B}\sum_{r'}\egorespentropy{\infty}{A}{B}{r'}
	\log_{\nummessages}\left(1+ \frac{\normalizedcoef{3}{B}
	\egorespentropy{\infty}{A}{B}{r'}}{\normalizedcoef{1}{B} 
	\responsedispcond{B}{\rprime}{r}+\normalizedcoef{4}{B}}\right)
\end{equation}
The right hand side is therefore independent of \(r\) and \(s\). It is thus constant and
corresponds to a uniform \(\egomemcond{\infty}{A}{B}{r}{s}\). Hence, 
using \eqnref{eq:asymptoticegoisselectprob}, the asymptotic selection probability is also uniform and the asymptotic marginal probability of 
messages~\(\genericprobmarg{\infty}{A}{r}\) is the uniform probability distribution, as 
described in \secref{sec:stationarystateequations}. 
Consistent with the choice of the coefficients, \(\genericprobmarg{\infty}{A}{r}\) has maximal entropy.

Next, note that if \(\expectationparam{A}\) is large (but strictly lower than one), 
convergence to the stationary state is mainly dominated by the ego memory. 
However, since the ego memory is based on actual message selections, the early steps of the interaction, when few messages have been exchanged and many 
of the \(\egomemjoint{t}{A}{B}{r}{s}\) are zero, are driven  by the response disposition and by the noise term. However, as soon as 
the stimulus-response pair  \(\transitionchoice{r}{s}\) has occurred once, the transition probability \(\selectprobcond{t}{A}{B}{r}{s}\) will
be dominated by the ego term   \(\selectprobcond{t}{A}{B}{r}{s}\) 
and the agent will tend to reproduce the previous answer.
The noise term allows the system to escape periodic cycles generated by the ego memory, 
but the time required to reach the asymptotic state can be very long. In practical terms, 
this means that 
when \(\randomparamtrunc\) is small and \(\expectationparamtrunc\) is large for each of 
the two agents, the interaction will
correspond to metastability, with limit cycles occurring on long times. Also, though there exists 
a unique asymptotic state as soon as \(\randomparamtrunc>0\), there may exist a large number of distinct metastable state.
Thus, when \(\expectationparamtrunc\) is large for both agents, we expect an 
effective (that is on the time scale of a typical interaction) ergodicity breaking 
with a wide variety of metastable states.

To summarize, the roles of the various coefficients in \eqnref{UXY} are:
\begin{itemize}

\item \(\knowledgeparamtrunc\) emphasizes the role of the response disposition. When \(\knowledgeparamtrunc\) is nearly one for both agents, the asymptotic behavior
is determined by the spectral properties of the product of the conditional response disposition.

\item \(\expectationparamtrunc\) enhances the tendency of an agent to reproduce its previous responses. It has a strong influence
on the transients and when \(\expectationparamtrunc\) is large,  many metastable states may be present.

\item \(\uncertaintyparamtrunc\) drives the agent to either pursue or avoid uncertainty. 
A positive \(\uncertaintyparamtrunc\)  favors
the increase of entropy, while a negative   \(\uncertaintyparamtrunc\) penalizes responses increasing the entropy.   

\item \(\randomparamtrunc\) is a noise term ensuring ergodicity, even 
though the characteristic time needed to reach stationarity (measured using, \eg, the spectral gap in the asymptotic selection probability matrices)
can be very long.
  
\end{itemize}


\bibliographystyle{plainnat}
\bibliography{references}

\clearpage


\begin{figure}[thbp]
	\centering
	\includegraphics[width=6cm,clip=false]{\figpath evol_dist_mode1_c1_1_c2_0_c3_0_c4_0}
	\caption{Evolution of the distance between the vector $\egomemmarg{t}{A}{r}$
	and the first mode of $\selectprobcondmatrix{\infty}{A}{B}\selectprobcondmatrix{\infty}{B}{A}$
	as the number of interaction sequences increases.}
	\label{FigConv}  
\end{figure}

\begin{figure}[thbp]
	\centering
	\includegraphics[width=10cm,clip=false]{\figpath Evolution_joint_entropy_moy_vs_t_nS11}
	\caption{Evolution of the joint entropy
	of alter and ego memory with infinite 
	($\altforgetparamabbrev=\egoforgetparamabbrev=1$) and finite
	($\altforgetparamabbrev=\egoforgetparamabbrev=0.99$) memories.}
	\label{FjointS}  
\end{figure}


\begin{figure}[thbp]
	\centering
	\includegraphics[width=10cm,clip=false]{\figpath Matrices_Asymptotiques_nS11_lambda0_99_moy10}
	\caption{Average asymptotic memories for \agent{A}
	after $128000$  steps, 
	with finite memories ($\altforgetparamabbrev=\egoforgetparamabbrev=0.99$).	
	The matrices shown here are calculated by averaging over $10$  initial conditions.
	The sizes of the red circles are proportional to the corresponding matrix elements, 
	while the sizes of the blue
	squares are proportional to the mean square deviations.}
	\label{FMatrAs4}  
\end{figure}

\end{document}

%% file: communicationmacros.tex
\newcommand{\figpath}{epsfigures/}

\newcommand{\booktitle}[1]{\textit{#1}}

\newcommand{\agentformat}[1]{\ensuremath{#1}}
\newcommand{\agent}[1]{agent~\agentformat{#1}}
\newcommand{\agentsxandy}[2]{agents~\agentformat{#1} and~\agentformat{#2}}

\newcommand{\nummessages}{\ensuremath{n_{S}}}
\newcommand{\egomemsymb}{\ensuremath{\mathcal{E}}}
\newcommand{\altmemsymb}{\ensuremath{\mathcal{A}}}
\newcommand{\respdispsymb}{\ensuremath{\mathcal{Q}}}
\newcommand{\selectprobsymb}{\ensuremath{\mathcal{U}}}
\newcommand{\respuncertsymb}{\ensuremath{\mathcal{H}}}
\newcommand{\normfactorsymb}{\ensuremath{\mathcal{N}}}
\newcommand{\genericprobsymb}{\ensuremath{\mathcal{P}}}
\newcommand{\normcoefsymb}{\ensuremath{\beta}}

\newcommand{\egomemmatrixsymb}{\ensuremath{\mathbf{E}}}
\newcommand{\altmemmatrixsymb}{\ensuremath{\mathbf{A}}}
\newcommand{\respdispmatrixsymb}{\ensuremath{\mathbf{Q}}}
\newcommand{\selectprobmatrixsymb}{\ensuremath{\mathbf{U}}}
\newcommand{\respuncertmatrixsymb}{\ensuremath{\mathbf{H}}}
\newcommand{\normunifmatrixsymb}{\ensuremath{\mathbf{S}}}
\newcommand{\genericprobmatrixsymb}{\ensuremath{\mathbf{P}}}

\newcommand{\timeind}[1]{\ensuremath{^{(#1)}}}
\newcommand{\margrvind}[1]{\ensuremath{_{#1}}}
\newcommand{\jointrvind}[2]{\ensuremath{_{#1 #2}}}
\newcommand{\condrvind}[2]{\ensuremath{_{#1 \given #2}}}

\newcommand{\egomemcondnoargs}[3]
	{\ensuremath{\egomemsymb\timeind{#1}\condrvind{#2}{#3}}}
\newcommand{\egomemcond}[5]
	{\ensuremath{\egomemcondnoargs{#1}{#2}{#3}\condargs{#4}{#5}}}
\newcommand{\egomemjointnoargs}[3]
	{\ensuremath{\egomemsymb\timeind{#1}\jointrvind{#2}{#3}}}
\newcommand{\egomemjoint}[5]
	{\ensuremath{\egomemjointnoargs{#1}{#2}{#3}\jointargs{#4}{#5}}}
\newcommand{\egomemmargnoargs}[2]
	{\ensuremath{\egomemsymb\timeind{#1}\margrvind{#2}}}
\newcommand{\egomemmarg}[3]
	{\ensuremath{\egomemmargnoargs{#1}{#2}\margargs{#3}}}

\newcommand{\altmemcondnoargs}[3]
	{\ensuremath{\altmemsymb\timeind{#1}\condrvind{#2}{#3}}}
\newcommand{\altmemcond}[5]
	{\ensuremath{\altmemcondnoargs{#1}{#2}{#3}\condargs{#4}{#5}}}
\newcommand{\altmemjointnoargs}[3]
	{\ensuremath{\altmemsymb\timeind{#1}\jointrvind{#2}{#3}}}
\newcommand{\altmemjoint}[5]
	{\ensuremath{\altmemjointnoargs{#1}{#2}{#3}\jointargs{#4}{#5}}}

\newcommand{\memoryentropy}[1]{\ensuremath{\respuncertsymb\left[#1\right]}}
\newcommand{\respentropy}[4]{\memoryentropy{\altmemcond{#1}{#2}{#3}{\cdot}{#4}}}
\newcommand{\egorespentropy}[4]{\memoryentropy{\egomemcond{#1}{#2}{#3}{\cdot}{#4}}}

\newcommand{\timedeprespdisp}[2]
	{\ensuremath{\respdispsymb\timeind{#1}_{#2}}}
\newcommand{\responsedispcondnoargs}[1]
	{\ensuremath{\respdispsymb_{#1}}}
\newcommand{\responsedispcond}[3]
	{\ensuremath{\responsedispcondnoargs{#1}\condargs{#2}{#3}}}
\newcommand{\responsedispjointnoargs}[1]
	{\ensuremath{\respdispsymb_{#1}}}
\newcommand{\responsedispjoint}[3]
	{\ensuremath{\responsedispjointnoargs{#1}\jointargs{#2}{#3}}}
\newcommand{\responsedispmargnoargs}[1]
	{\ensuremath{\respdispsymb_{#1}}}
\newcommand{\responsedispmarg}[2]
	{\ensuremath{\responsedispmargnoargs{#1}\margargs{#2}}}
\newcommand{\responsedispcondabbrev}[1]{\ensuremath{\respdispsymb}}

\newcommand{\selectprobcondnoargs}[3]
	{\ensuremath{\selectprobsymb\timeind{#1}\condrvind{#2}{#3}}}
\newcommand{\selectprobcond}[5]
	{\ensuremath{\selectprobcondnoargs{#1}{#2}{#3}\condargs{#4}{#5}}}

\newcommand{\selectprobcondtrunc}[1]{\ensuremath{\selectprobsymb\timeind{#1}}}

\newcommand{\genericprobjointnoargs}[3]
	{\ensuremath{\genericprobsymb\timeind{#1}\jointrvind{#2}{#3}}}
\newcommand{\genericprobjoint}[5]
	{\ensuremath{\genericprobjointnoargs{#1}{#2}{#3}\jointargs{#4}{#5}}}
\newcommand{\genericprobmargnoargs}[2]
	{\ensuremath{\genericprobsymb\timeind{#1}\margrvind{#2}}}
\newcommand{\genericprobmarg}[3]
	{\ensuremath{\genericprobmargnoargs{#1}{#2}\margargs#3}}
\newcommand{\genericprobmargabbrev}[1]{\ensuremath{\genericprobsymb\timeind{#1}}}

\newcommand{\standardmatrixmemoryformat}[4]
	{\ensuremath{#1\timeind{#2}\condrvind{#3}{#4}}}

\newcommand{\normalizeduniformprobmatrix}{\ensuremath{\normunifmatrixsymb}}

\newcommand{\egomemcondmatrix}[3]
	{\standardmatrixmemoryformat{\egomemmatrixsymb}{#1}{#2}{#3}}
\newcommand{\responsedispcondmatrix}[1]
	{\ensuremath{\respdispmatrixsymb_{#1}}}
\newcommand{\selectprobcondmatrix}[3]
	{\standardmatrixmemoryformat{\selectprobmatrixsymb}{#1}{#2}{#3}}

\newcommand{\selectprobcondmatrixtrunc}[1]
	{\ensuremath{\selectprobmatrixsymb\timeind{#1}}}
\newcommand{\genericprobmargmatrix}[2]
	{\ensuremath{\genericprobmatrixsymb\timeind{#1}\margrvind{#2}}}
\newcommand{\genericprobmargmatrixtrunc}[1]
	{\ensuremath{\genericprobmatrixsymb\timeind{#1}}}

\newcommand{\respentropymatrix}[3]
	{\ensuremath{\respuncertmatrixsymb \left [
	 \standardmatrixmemoryformat{\altmemmatrixsymb}{#1}{#2}{#3} \right ]}}

\newcommand{\uniformmatrixeigenvector}{\ensuremath{\mathbf{u}}}
\newcommand{\respentropymatrixeigenvectortrunc}[1]
	{\ensuremath{\mathbf{v}\timeind{#1}}}
\newcommand{\respentropymatrixeigenvector}[2]
	{\ensuremath{\mathbf{v}_{#2}\timeind{#1}}}

\newcommand{\forgetparamsymbol}{\lambda}
\newcommand{\generalforgetparam}[2]{\forgetparamsymbol_{#1}^{(#2)}}
\newcommand{\generalforgetparamabbrev}[1]{\generalforgetparam{}{#1}}
\newcommand{\altforgetparam}[1]{\generalforgetparam{#1}{\altmemsymb}}
\newcommand{\altforgetparamabbrev}{\generalforgetparamabbrev{\altmemsymb}}
\newcommand{\egoforgetparam}[1]{\generalforgetparam{#1}{\egomemsymb}}
\newcommand{\egoforgetparamabbrev}{\generalforgetparamabbrev{\egomemsymb}}

\newcommand{\genericcoef}[2]{\ensuremath{c_{#1}^{(#2)}}}

\newcommand{\genericcoeftrunc}[1]{\ensuremath{c_{#1}}}
\newcommand{\normalizedcoef}[2]{\ensuremath{\normcoefsymb_{#1}^{(#2)}}}

\newcommand{\knowledgeparam}[1]{\genericcoef{1}{#1}}
\newcommand{\expectationparam}[1]{\genericcoef{2}{#1}}
\newcommand{\uncertaintyparam}[1]{\genericcoef{3}{#1}}
\newcommand{\randomparam}[1]{\genericcoef{4}{#1}}

\newcommand{\knowledgeparamtrunc}{\genericcoeftrunc{1}}
\newcommand{\expectationparamtrunc}{\genericcoeftrunc{2}}
\newcommand{\uncertaintyparamtrunc}{\genericcoeftrunc{3}}
\newcommand{\randomparamtrunc}{\genericcoeftrunc{4}}

\newcommand{\affinitysymbol}{\ensuremath{\mathcal{G}}}
\newcommand{\affinity}[2]{\ensuremath{\affinitysymbol_{#1#2}}}
\newcommand{\affinityprod}[2]{\affinity{#1}{#2}\affinity{#2}{#1}}
\newcommand{\affinityconst}{\ensuremath{k}}

\newcommand{\normfactorcond}[3]
	{\ensuremath{\normfactorsymb}\timeind{#1}\condrvind{#2}{#3}}
\newcommand{\normfactorcondabbrev}[3]
	{\ensuremath{\normfactorsymb}\timeind{#1}}

\newcommand{\messagechoicesymbol}{\ensuremath{m}}
\newcommand{\messagechoice}[1]{\ensuremath{\messagechoicesymbol_{#1}}}

\newcommand{\messagesequence}[4]{\ensuremath{\messagechoice{#1}, \messagechoice{#2},
							\ldots, \messagechoice{#3}, \messagechoice{#4}}}
\newcommand{\infinitemessagesequence}[4]
	{\ensuremath{\messagesequence{#1}{#2}{#3}{#4}, \ldots}}

\newcommand{\transitionchoice}[2]{\ensuremath{\left(#1, #2\right)}}
\newcommand{\namedinfinitemessagesequence}{\ensuremath{\tilde{\messagechoicesymbol}}}
\newcommand{\messagechoicepair}[2]
	{\ensuremath{(\messagechoice{#1}, \messagechoice{#2})}}

\newcommand{\seqempfreqsymb}{\ensuremath{\mathcal{F}}}
\newcommand{\seqempfreq}[1]{\seqempfreqsymb^{#1}}

\newcommand{\seqdistancesymb}{\ensuremath{d}}
\newcommand{\seqdistance}[2]{\ensuremath{\seqdistancesymb\left(#1, #2\right)}}
\newcommand{\sqseqdistance}[2]{\ensuremath{\seqdistancesymb^{2}\left(#1, #2\right)}}
\newcommand{\threshsymb}{\ensuremath{\eta}}

\newcommand{\seqdistancethresh}[1]{\threshsymb_{#1}}
\newcommand{\sequpdaterate}[1]{\ensuremath{k_{#1}}}

\newcommand{\apriori}{\textit{a priori}}

\newcommand{\ie}{i.e.}
\newcommand{\etc}{etc.} 
\newcommand{\eg}{e.g.}

\newcommand{\eqnrefformat}[1]{(\ref{#1})}
\newcommand{\eqnref}[1]{Eq.~\eqnrefformat{#1}}
\newcommand{\eqnxandy}[2]{Eqs.~\eqnrefformat{#1} and~\eqnrefformat{#2}}

\newcommand{\figrefformat}[1]{\ref{#1}}
\newcommand{\figref}[1]{Fig.~\figrefformat{#1}}

\newcommand{\secrefformat}[1]{\ref{#1}}
\newcommand{\secref}[1]{section~\secrefformat{#1}}
\newcommand{\appendref}[1]{section~\secrefformat{#1}}

\newcommand{\half}{\frac{1}{2}}

\newcommand{\krondelta}[2]{\delta^{#1}_{#2}}
\newcommand{\rprime}{\ensuremath{r^{\prime}}}

\newcommand{\entropybasen}[3]{-\sum_{#2} #1 \log_{#3} #1}

\newcommand{\set}[1]{\ensuremath{\left\{#1\right\}}}
\newcommand{\given}{\mid}

\newcommand{\jointargs}[2]{\ensuremath{\left(#1, #2\right)}}
\newcommand{\condargs}[2]{\ensuremath{\left(#1 \given #2\right)}}
\newcommand{\margargs}[1]{\ensuremath{\left(#1\right)}}

\newcommand{\mathcomma}{\quad,}
\newcommand{\mathperiod}{\quad.}
